\documentclass[twocolumn]{aastex63}
\usepackage{booktabs}
\usepackage{natbib}
\shorttitle{Evolution of a peculiar type Ibn supernova SN~2019wep}
\shortauthors{Gangopadhyay et al.}
\begin{document}

\title{Evolution of A Peculiar Type Ibn Supernova SN~2019wep}

\correspondingauthor{Anjasha Gangopadhyay}
\email{anjashagangopadhyay@gmail.com}

\author[0000-0002-3884-5637]{Anjasha Gangopadhyay}
\affiliation{Hiroshima Astrophysical Science Centre, Hiroshima University, 1-3-1 Kagamiyama, Higashi-Hiroshima, Hiroshima 739-8526, Japan}
\affiliation{Aryabhatta Research Institute of observational sciencES, Manora Peak, Nainital 263 002 India}

\author[0000-0003-1637-267X]{Kuntal Misra}
\affiliation{Aryabhatta Research Institute of observational sciencES, Manora Peak, Nainital 263 002 India}

\author[0000-0002-0832-2974]{Griffin Hosseinzadeh}
\affiliation{Steward Observatory, University of Arizona, 933 North Cherry Avenue, Tucson, AZ 85721, USA}

\author[0000-0001-7090-4898]{Iair Arcavi}
\affiliation{The School of Physics and Astronomy, Tel Aviv University, Tel Aviv 69978, Israel}
\affiliation{CIFAR Azrieli Global Scholars program, CIFAR, Toronto, Canada}

\author{Craig Pellegrino}
\affiliation{Las Cumbres Observatory, 6740 Cortona Drive Suite 102, Goleta, CA, 93117-5575 USA}
\affiliation{Department of Physics, University of California, Santa Barbara, CA 93106-9530, USA}
\author[0000-0002-7334-2357]{Xiaofeng Wang}
\affiliation{Physics Department and Tsinghua Center for Astrophysics, Tsinghua University, China}
\affiliation{Beijing Planetarium, Beijing Academy of Sciences, Beijing, 100044}

\author{D. Andrew Howell}
\affiliation{Las Cumbres Observatory, 6740 Cortona Drive Suite 102, Goleta, CA, 93117-5575 USA}
\affiliation{Department of Physics, University of California, Santa Barbara, CA 93106-9530, USA}

\author{Jamison Burke}
\affiliation{Las Cumbres Observatory, 6740 Cortona Drive Suite 102, Goleta, CA, 93117-5575 USA}
\affiliation{Department of Physics, University of California, Santa Barbara, CA 93106-9530, USA}

\author{Jujia Zhang}
\affiliation{Yunnan Astronomical Observatory of China, Chinese Academy of Sciences, Kunming, 650216, China}
\affiliation{Key Laboratory for the Structure and Evolution of Celestial Objects,CAS, Kunming 650216, China}

\author{Koji Kawabata}
\affiliation{Hiroshima Astrophysical Science Centre, Hiroshima University, 1-3-1 Kagamiyama, Higashi-Hiroshima, Hiroshima 739-8526, Japan}

\author{Mridweeka Singh}
\affiliation{Indian Institute of Astrophysics, 2nd Block, 100 Feet Rd, Koramangala, Bengaluru, Karnataka, India 560034}
\affiliation{Korea Astronomy and Space Science Institute, 776 Daedeokdae-ro, Yuseong-gu, Daejeon 34055, Republic of Korea}

\author[0000-0001-6191-7160]{Raya Dastidar}
\affiliation{Millennium Institute of Astrophysics, Nuncio Monsenor Sótero Sanz 100, Providencia, Santiago, 8320000 Chile}

%\author{Beijing Planetarium}
%\affiliation{Beijing Academy of Sciences, Beijing, 100044, China}

\author[0000-0002-1125-9187]{Daichi Hiramatsu}
\affiliation{Las Cumbres Observatory, 6740 Cortona Drive Suite 102, Goleta, CA, 93117-5575 USA}
\affiliation{Department of Physics, University of California, Santa Barbara, CA 93106-9530, USA}

\author{Curtis McCully}
\affiliation{Las Cumbres Observatory, 6740 Cortona Drive Suite 102, Goleta, CA, 93117-5575 USA}
\affiliation{Department of Physics, University of California, Santa Barbara, CA 93106-9530, USA}

\author{Jun Mo}
\affiliation{Physics Department and Tsinghua Center for Astrophysics, Tsinghua University, China}

\author{Zhihao Chen}
\affiliation{Physics Department and Tsinghua Center for Astrophysics, Tsinghua University, China}

\author[0000-0002-1089-1519]{Danfeng Xiang} 
\affiliation{Physics Department and Tsinghua Center for Astrophysics, Tsinghua University, China}



\begin{abstract}

We present a high-cadence short term photometric and spectroscopic monitoring campaign of a type Ibn SN~2019wep, which is one of the rare SN~Ibn after SNe~2010al and 2019uo to display signatures of flash ionization (\ion{He}{2}, \ion{C}{3}, \ion{N}{3}). We compare the decline rates and rise time of SN 2019wep with other SNe~Ibn and fast transients. The post-peak decline in all bands (0.1 mag d$^{-1}$) are consistent with SNe~Ibn but less than the fast transients. On the other hand, the $\Delta$m$_{15}$ values are slightly lower than the average values for SNe~Ibn but consistent with the fast transients. The rise time is typically shorter than SNe~Ibn but longer than fast transients. SN~2019wep lies at the fainter end of SNe~Ibn but possesses an average luminosity amongst the fast transients sample. The peculiar color evolution places it between SNe~Ib and the most extreme SNe~Ibn. The bolometric light curve modelling shows resemblance with SN~2019uo with ejecta masses consistent with SNe~Ib. SN~2019wep belongs to the ``P~cygni" sub-class of SNe~Ibn and shows faster evolution in line velocities as compared to the ``emission" sub-class. The post-maximum spectra show close resemblance with ASASSN-15ed hinting it to be of SN~Ib nature. The low \ion{He}{1} CSM velocities and residual H$\alpha$ further justifies it and gives evidence of an intermittent progenitor between WR and LBV star.

\end{abstract}


\section{Introduction}
\label{1}
Type Ibn Supernovae (hereafter SNe~Ibn) are a rare class of core-collapse SNe (CCSNe) undergoing interaction with hydrogen poor circumstellar medium (CSM). Interaction, in general, produces narrow emission lines --- broader than \ion{H}{2} regions but narrower than the lines arising from the outer ejecta of the SN \citep{2007Natur.447..829P}. However, in some cases interaction happens below the photosphere without any observable narrow emission lines \citep[e.g.][]{2017ApJ...838...28M,2018MNRAS.477...74A}. SNe~Ibn are characterised by narrow lines of He ($\sim$2000~km~s$^{-1}$) in their spectra and was introduced with the discovery of SN~2006jc \citep{2007Natur.447..829P}. This is defined in analogy with SNe~IIn, which show narrow H features \citep{1990MNRAS.244..269S}. SNe that are embedded in dense CSM may also show short-lived narrow high ionization emission lines (at $\leq$10~days) owing to the recombination of the CSM following the shock breakout flash or through interaction of the SN ejecta with nearby CSM. These features are known as ``flash features" \citep[e.g.][]{2014Natur.509..471G,2019ApJ...882..102G}. 

Unlike the light curves of SNe~IIn which show great photometric diversity \citep{2020A&A...637A..73N}, SNe~Ibn light curves (though small in number as only less than 40 SNe~Ibn are known till date \citep{2019ApJ...871L...9H}) are surprisingly homogeneous with a uniform light curve shape and decay rates of 0.05--0.15~mag~d$^{-1}$ \citep{2017ApJ...836..158H}. They typically rise within a timescale of $\leq$ 15 days and reach a peak absolute magnitude of M$_{R}$ = $-$ 19 mag \citep{2017ApJ...836..158H}, with exceptions like ASASSN-14ms which has a peak luminosity of M$_{V}$ $\sim-$20.5 mag, lying between normal SNe~Ibn and superluminous SNe \citep{2021ApJ...917...97W}. \cite{2016MNRAS.456..853P}, however, highlighted the existence of many outliers in the SN~Ibn group which are slow decliners, for example, OGLE-2012-SN-006 \citep{2015MNRAS.449.1941P} and SN~2010al \citep{2015MNRAS.449.1921P}. The fast-rising and decaying nature of SNe~Ibn enables their comparison with the fast transients (\citet{2014ApJ...794...23D, 2016ApJ...819...35A, 2021arXiv211015370P}) which may hint toward their similar origin. The light curves of Stripped envelope (SESNe; typically Ibs) are in general powered by radioactivity but in the case of SNe~Ibn the light curves are too steep to reconcile for the amount of $^{56}$Ni contributing to the radioactive luminosity \citep{2016ApJ...824..100M}. Instead, a combination of Ni-decay and CSM interaction are required to reproduce the light curve \citep{2019arXiv191006016K,2019arXiv190512623W,2020ApJ...889..170G}. 

In the past few years, one of the most unsettling question has been regarding the progenitors giving rise to SNe~Ibn explosions. The homogeneity of the light curves does not necessarily imply the homogeneity of the progenitors, which are largely dependent on the CSM configuration. Unlike SNe~IIn where some have been linked to H-rich luminous blue variable (LBV) stars (e.g., \citet{2009Natur.458..865G}; \citet{2010ApJ...709..856S}), there have been no direct detection of SNe~Ibn progenitors to date. There are two precursor detections of massive SN Ibn progenitors -- the luminous outburst of the prototypical SN~Ibn 2006jc, observed two years before the explosion \citep{2007ApJ...657L.105F,2007Natur.447..829P,2008ApJ...680..568S} and the other for SN~2019uo where precursor starts $\sim$ 340 days before the explosion and is observed over 35 days \citep{2021ApJ...907...99S}. The progenitor of SN 2006jc is believed to be a massive star with remaining LBV properties \citep{2007Natur.447..829P}. The stripping of the outer envelope can occur also from the massive binaries \citep{2007ApJ...657L.105F}. \cite{2021ApJ...907...99S} suggests that the progenitors could be Wolf-Rayet (WR) stars that have shed their hydrogen envelopes or massive stars that are stripped by a binary partner. Recent evidence of a low mass progenitor host has been proposed for PS1-12sk, which occurred in a non-star forming galaxy \citep{2013ApJ...769...39S,2019ApJ...871L...9H}. Based on the late time UV/Optical {\it HST} images, an interacting binary progenitor scenario was inferred for SNe~2006jc and 2015G \citep{2020MNRAS.491.6000S}.\\

\noindent
{\bf Discovery:}

SNe~Ibn 2019wep was discovered on 2019 December 07.96979 UT by Xing Gao \citep{2019TNSCR2565....1Z} in Xingming Observatory Sky Survey (C42) images at an unfiltered mag of $\sim$ 18.2 which was estimated by several 40 sec survey images. The SN was located at $\mathrm{R.A.} = 11^\mathrm{h}04^\mathrm{m}36.99^\mathrm{s}$, $\mathrm{Dec} = +45\degr58'39".5$ (equinox J2000.0) \citep{2019TNSTR2556....1M} which is about 2\farcs6 east and 27\farcs6 south of the center of the host galaxy PGC33464.
\cite{2019ATel13337....1Z} obtained the classification spectrum of SN 2019wep with the 2.4 m Li-Jiang Telescope (LJT+YFOSC) of Yunnan Observatories, China. The spectrum depicted a blue continuum and highly ionized ``flash features" of \ion{He}{2}, \ion{H}{1} and \ion{C}{4}. This spectrum matches with that of a young SN~II with signatures of interaction. Another spectrum taken a few days later shows evidence of narrow P~cygni \ion{He}{1} features classifying it as a SN~Ibn. SN~2019wep is one of the rare SNe~Ibn to show signatures of flash ionisation after SNe~2019uo and 2010al.\\

\noindent
{\bf Estimation of explosion epoch:}

\cite{2019TNSTR2556....1M} reported the non-detection of a source in archival images of Xingming Observatory Sky Survey (C42) on 2019-05-30 (limiting mag 19.0) while the Zwicky Transient Facility (ZTF\footnote{\url{https://www.wis-tns.org/object/2019wep};  \citealt{2014htu..conf...27B, 2019PASP..131f8003B}}) reported a last non-detection of the source on 2019-11-25 (MJD 58812) in $r$-band at a limiting magnitude of 19.89 mag. Four pre-maximum unfiltered data points of SN~2019wep are available in TNS\footnote{\url{https://www.wis-tns.org/object/2019wep}} which match well with our $V$-band data. We combine the unfiltered magnitudes with our $V$-band data (since the passbands are close) and perform a parabolic fit on the combined light curve of SN 2019wep. The magnitudes are converted to flux and the fit is performed using the data up to $\sim$ 6 days after the discovery. The best-fit coefficients are used to find the roots of the equation, i.e., the value of time for which the flux equals zero. This gives the explosion epoch to be MJD 58824.5 $\pm$ 2. The error is estimated from the fitting error associated with the data.\\

\noindent
{\bf Distance and Extinction:}

Adopting $H_0 = 73$~km~s$^{-1}$~Mpc$^{-1}$, $\Omega_{matter}$=0.27 and $\Omega_{vaccum}$=0.73, we obtain a luminosity distance of 108.3~Mpc (corresponding to a redshift $z=$0.025 \footnote{\url{https://ned.ipac.caltech.edu/byname?objname=PGC33464&hconst=73&omegam=0.27&omegav=0.73&corr_z=4}}) for SN~2019wep. The Milky Way extinction along the line of sight of SN~2019wep is $A_V = 0.031$~mag \citep{2011ApJ...737..103S}. For estimating the extinction due to the host galaxy, we estimate equivalent widths of the \ion{Na}{1}D line in the three spectra of SN~2019wep taken on 2019-12-16, 2019-12-19 and 2019-12-24. Using the formulation by \cite{1997A&A...318..269M} and \cite{2012MNRAS.426.1465P}, we estimate host galaxy $A_V = 0.0328$~mag. The estimated extinction results in matching the $(B-V)$ colors of SN~2019wep in close agreement with that of SNe~2010al and 2019uo. Thus, we adopt a total $A_V = 0.0638$~mag. We use these values of distance and extinction throughout the paper. 

In this paper we aim to investigate the photometric and spectroscopic evolution of the peculiar SN~Ibn 2019wep which not only showed signatures of flash ionisation but also shows some similarities with SN~Ib at late phases. In Section \ref{2}, we discuss the procedure of photometric and spectroscopic data reduction. In Section \ref{3}, the photometric evolution of SN~2019wep is studied and compared with other SNe~Ibn and fast transients. In Section \ref{4}, we discuss the interpretation of the observed spectral features including the late-time distinct P~cygni features and unique H$\alpha$ behavior.  Our conclusions are given in Section \ref{5}.

\section{Data Acquisition and Reduction}
\label{2}

We observed SN~2019wep with the 2m Las Cumbres Observatory (LCO) telescopes in the \textit{UBVgri} filters from $\sim$3 to 98 days after explosion under the Global Supernovae Project (GSP) program. The photometric observations in UBVRI/ugri were also taken with the 0.8m Tsinghua-NAOC Telescope (TNT; \citealt{2012RAA....12.1585H}), Xinglong Observatory, China. Image subtraction was done using High Order Transform of PSF ANd Template Subtraction  \footnote{\url{https://github.com/acbecker/hotpants}}. The instrumental magnitudes were estimated using  DAOPHOT\footnote{Dominion Astrophysical Observatory Photometry}. The LCO image subtraction and photometry were done with \texttt{lcogtsnpipe}\footnote{\url{https://github.com/svalenti/lcogtsnpipe}} (see \citealp{2011MNRAS.416.3138V,2016MNRAS.459.3939V}). The instrumental SN magnitudes in {\it gri} bands and {\it UBV} bands were calibrated with respect to the Sloan Digital Sky Survey (SDSS) catalog and Landolt standard fields respectively. 

The photometry of SN~2019wep is presented in Table \ref{tab:observation_log_2019wep}.

% \begin{center}
%\begin{longrotatetable}
%\begin{longtable}{|c | c | c | c | c | c | c | c|}
\begin{table*}
\centering
% \footnotesize\addtolength{\tabcolsep}{-2pt}
\caption{Photometry of SN~2019wep. The phase is measured with respect to $V$-max (MJD$_{max}$=58828.5).} 
\resizebox{\hsize}{!}{\begin{tabular}{c c c c c c c c c c c}
\toprule
Date  & JD & Phase        & $U$     & $B$             & $g$              & $V$                 & $r$              &  $i$         & Telescope\\    
(yyyy-mm-dd) & (2400000+)   & (day)    & (mag)   & (mag)           & (mag)            & (mag)               & (mag)            & (mag)        &            \\
\midrule
2019-12-12 & 58829.35  &  0.85 & ---	           &    17.19$\pm$0.03  &  17.11$\pm$0.03  &   17.29$\pm$0.03   &  17.42$\pm$0.10   &       ---                &            LCO \\
2019-12-12 & 58829.35  &  0.85 & ---		   &    17.20$\pm$0.03  &  17.18$\pm$0.04  &   17.38$\pm$0.04   &  ---              &   ---                    &            LCO \\
2019-12-13 & 58830.41  &  1.91 & 16.49$\pm$0.03	   &    17.38$\pm$0.02  &  17.35$\pm$0.03  &   17.41$\pm$0.03   &  17.48$\pm$0.03   &        17.91$\pm$0.07    &     LCO \\
2019-12-13 & 58830.41  &  1.91 & 16.45$\pm$0.04    &    17.35$\pm$0.03  &  17.28$\pm$0.02  &   17.43$\pm$0.03   &   17.57$\pm$0.03  &       17.75$\pm$0.07     &           LCO \\
2019-12-15 & 58832.36  & 3.86 & 16.86$\pm$0.04	   &    17.48$\pm$0.03  &  17.36$\pm$0.03  &   17.59$\pm$0.04   &   17.62$\pm$0.03  &       17.68$\pm$0.05     &     LCO \\
2019-12-15 & 58832.37  &  3.86 & 16.85$\pm$0.05	   &    17.56$\pm$0.03  &  17.28$\pm$0.03  &    ---             &   17.59$\pm$0.03  &       17.63$\pm$0.05     &     LCO \\
2019-12-18 & 58835.35  &  6.85 & 17.33$\pm$0.05   &    18.05$\pm$0.03  &  17.84$\pm$0.02  &   17.93$\pm$0.03   &   17.88$\pm$0.03  &       18.02$\pm$0.03     &           LCO \\
2019-12-18 & 58835.35  &  6.85 &  ---		   &    18.06$\pm$0.03  &  17.83$\pm$0.02  &   17.93$\pm$0.04   &   17.86$\pm$0.03  &       18.05$\pm$0.03     &     LCO \\
2019-12-18 & 58835.73  &  7.25 & ---   &    18.34$\pm$0.03  &  18.18$\pm$0.03  &   17.93$\pm$0.05   &   18.06$\pm$0.02  &       18.05$\pm$0.03     &           TNT \\
2019-12-19 & 58836.36  &  7.86 &  17.63$\pm$0.03  &    18.10$\pm$0.02  &  17.94$\pm$0.01  &   17.97$\pm$0.03   &   17.89$\pm$0.02  &       17.97$\pm$0.03     &         LCO \\
2019-12-19 & 58836.37  &  7.87 &  17.49$\pm$0.03  &    18.13$\pm$0.02  &  17.95$\pm$0.01  &   17.98$\pm$0.03   &   17.88$\pm$0.02  &       18.07$\pm$0.03     &         LCO \\
2019-12-21 & 58838.33  &  9.83 &  17.74$\pm$0.04  &    18.43$\pm$0.03  &  18.13$\pm$0.01  &   18.16$\pm$0.03   &   18.03$\pm$0.02  &       18.12$\pm$0.03     &  LCO \\
2019-12-21 & 58838.34  &  9.84 &  17.80$\pm$0.04  &    18.35$\pm$0.02  &  18.14$\pm$0.01  &   18.11$\pm$0.03   &   18.06$\pm$0.02  &       18.18$\pm$0.04     &        LCO \\
2019-12-22 & 58839.47  &  10.97 &  18.01$\pm$0.03  &    18.64$\pm$0.02  &  18.33$\pm$0.01  &   18.27$\pm$0.02   &   18.14$\pm$0.02  &       18.23$\pm$0.03     &  LCO \\
2019-12-22 & 58839.48  &  10.98 &  18.02$\pm$0.03  &    18.62$\pm$0.02  &  18.36$\pm$0.01  &   18.31$\pm$0.02   &   ---             &       18.23$\pm$0.02     &  LCO \\
2019-12-24 & 58841.44  &  13.04 &  18.45$\pm$0.10  &    18.75$\pm$0.03  &  18.55$\pm$0.04  &   18.44$\pm$0.06   &   18.27$\pm$0.04  &       18.21$\pm$0.07     &        LCO \\   
2019-12-24 & 58841.44  &  13.04 &  18.37$\pm$0.08  &    ---             &  18.60$\pm$0.03  &   18.34$\pm$0.05   &   18.19$\pm$0.04  &       18.33$\pm$0.08     &         LCO \\
2019-12-27 & 58844.37  &  15.87 &  18.98$\pm$0.06  &   19.27$\pm$0.04   &  18.98$\pm$0.02  &   18.67$\pm$0.03   &   18.56$\pm$0.03  &       18.55$\pm$0.04     &         LCO \\
2019-12-27 & 58844.37  &  15.87 &  19.01$\pm$0.06  &   19.29$\pm$0.04   &  18.99$\pm$0.02  &   18.70$\pm$0.03   &   18.49$\pm$0.02  &       18.59$\pm$0.04     &  LCO \\
2019-12-27 & 58844.78  &  16.28 &  ---  &   19.30$\pm$0.09   &  19.00$\pm$0.05  &   18.71$\pm$0.04   &   18.50$\pm$0.03  &       18.51$\pm$0.03     &  TNT \\
2019-12-28 & 58845.80  &  17.30 &  ---  &   19.37$\pm$0.03   &  19.11$\pm$0.02  &   18.77$\pm$0.03   &   18.61$\pm$0.02  &       18.52$\pm$0.09     &  TNT \\
2019-12-29 & 58846.81  &  18.31 &  ---  &   19.47$\pm$0.05   &  19.28$\pm$0.03  &   18.86$\pm$0.06   &   18.81$\pm$0.06  &       18.91$\pm$0.09     &  TNT \\
2019-12-30 & 58847.78  &  19.28 &  ---  &   19.57$\pm$0.05   &  19.47$\pm$0.07  &   19.05$\pm$0.09   &   18.95$\pm$0.09  &       ---    &  TNT \\
2020-01-01  & 58849.82  & 21.32 &  ---  &   20.07$\pm$0.09   &  19.87$\pm$0.07  &   19.07$\pm$0.05   &   19.10$\pm$0.09  &       18.86$\pm$0.05     &  TNT \\
2020-01-02 & 58850.47  &  21.97 &  20.32$\pm$0.24  &   20.42$\pm$0.08   &  20.10$\pm$0.05  &   19.65$\pm$0.08   &   19.31$\pm$0.07  &       19.49$\pm$0.13     &        LCO \\
2020-01-02 & 58850.47  &  21.97 &  20.12$\pm$0.23  &   20.43$\pm$0.10   &  20.09$\pm$0.05  &   19.64$\pm$0.07   &   19.29$\pm$0.06  &       ---                &  LCO \\
2020-01-02 & 58850.75  &  22.25 &  ---  &   20.57$\pm$0.99   &  18.89$\pm$0.05  &   19.48$\pm$0.05   &   19.18$\pm$0.04  &       19.29$\pm$0.04     &  TNT \\
2020-01-03  & 58851.78  & 23.28 &  ---  &   20.39$\pm$0.08   &  20.09$\pm$0.05  &   19.59$\pm$0.05   &   19.29$\pm$0.03  &       18.97$\pm$0.04                &  TNT \\
2020-01-04  & 58852.76  & 24.26 &  ---  &   20.57$\pm$0.99   &  20.09$\pm$0.05  &   19.59$\pm$0.05   &   19.40$\pm$0.04  &       19.11$\pm$0.07               &  TNT \\
2020-01-05 & 58853.30  &  24.80 &  20.04$\pm$0.27  &   20.68$\pm$0.22   &  20.52$\pm$0.14  &   19.86$\pm$0.12   &   19.60$\pm$0.09  &       19.38$\pm$0.09     &  LCO \\
2020-01-05 & 58853.30  &  24.80 &  20.40$\pm$0.32  &   ---              &    ---           &   19.89$\pm$0.13   &   19.51$\pm$0.09  &       19.35$\pm$0.08     &        LCO \\
2020-01-06 & 58854.28  &  25.78 &  ---		   &   20.62$\pm$0.22   & 20.65$\pm$0.19   &   20.06$\pm$0.14   &   19.48$\pm$0.10  &       19.48$\pm$0.24     &         LCO \\
2020-01-06 & 58854.28  &  25.78 &  ---		   &   20.41$\pm$0.22   & ---              &   19.95$\pm$0.13   &   19.41$\pm$0.08  &       20.53$\pm$0.15     &         LCO \\
2020-01-14 & 58862.44  &  33.94 & --- 		   &   20.73$\pm$0.24   & ---              &   20.63$\pm$0.20   &   19.47$\pm$0.16  &       ---                &  LCO \\
2020-01-14 & 58862.46  &  33.94 & ---              &    ---             &  ---             &   20.32$\pm$0.17   &   19.96$\pm$0.21  &       ---                &        LCO \\
2020-01-18 & 58866.44  &  37.94 & --- 		   &   21.49$\pm$0.24   & 21.51$\pm$0.17   &   20.63$\pm$0.16   &   20.01$\pm$0.09  &        19.88$\pm$0.11    &  LCO \\
2020-01-18 & 58866.44  &  37.94 & --- 		   &   21.28$\pm$0.21   & 21.10$\pm$0.12   &   20.53$\pm$0.13   &   20.08$\pm$0.09  &        19.88$\pm$0.11    &  LCO \\
2020-01-24 & 58872.48  &  43.98 & --- 		   &   21.34$\pm$0.18   & 21.27$\pm$0.10   &   21.28$\pm$0.14   &   20.15$\pm$0.08  &        19.91$\pm$0.11    &        LCO \\
2020-01-24 & 58872.49  &  43.99 & --- 		   &   21.48$\pm$0.25   & 21.28$\pm$0.11   &   20.89$\pm$0.12   &   20.18$\pm$0.07  &        19.94$\pm$0.11    &  LCO \\
2020-01-28 & 58876.43  &  47.93 & --- 		   &   21.66$\pm$0.33   & 21.47$\pm$0.16   &   20.79$\pm$0.16   &   20.34$\pm$0.12  &        20.31$\pm$0.16    &         LCO \\
2020-01-28 & 58876.44  &  47.94 & --- 		   &   21.84$\pm$0.28   & 21.66$\pm$0.24   &   20.77$\pm$0.28   &   20.29$\pm$0.08  &         ---              &         LCO \\
2020-01-28 & 58876.48  &  47.98 & --- 		   &   21.80$\pm$0.33   & ---              &   20.76$\pm$0.13   &   ---             &      ---                 &  LCO \\
2020-01-28 & 58876.48  &  47.98 & --- 		   &   21.37$\pm$0.23   & ---              &   20.84$\pm$0.17   &   ---             &      ---                 &        LCO \\
2020-02-01 & 58880.47  &  51.97 &  		   &   21.39$\pm$0.24   & 21.43$\pm$0.12   &   20.66$\pm$0.12   &   20.39$\pm$0.10  &        20.20$\pm$0.13    &  LCO \\
2020-02-01 & 58880.48  &  51.98 & --- 		   &   21.52$\pm$0.24   & 21.56$\pm$0.16   &   20.81$\pm$0.16   &   20.42$\pm$0.11  &        20.33$\pm$0.15    &  LCO \\
2020-02-08 & 58887.41  &  58.91 & ---		   &   21.51$\pm$0.53   & 21.34$\pm$0.28   &   21.21$\pm$0.49   &   20.34$\pm$0.19  &        20.38$\pm$0.25    &        LCO \\
2020-02-08 & 58887.41  &  58.93 & ---              &   ---              & 21.88$\pm$0.42   &   20.67$\pm$0.34   &   21.57$\pm$0.44  &         21.76$\pm$0.24   &      LCO \\
 2020-03-15      & 58923.61  & 95.11 &  ---    & 22.67$\pm$0.39 & 22.27$\pm$0.31 & 22.05$\pm$0.45 & 21.99$\pm$0.43 & 21.88$\pm$0.13 & TNT \\
\bottomrule                                                                                        
%\multicolumn{4}{@{}l}{$^\dagger$ with respect to $\mathrm{MJD_{max}} = 58828.5$.}               
\end{tabular}}
\label{tab:observation_log_2019wep}     
% \newline
% $^\dagger$ with respect to $\mathrm{JD_{max}} = 2458508.65$.
% \newline
\end{table*}
%\end{longrotatetable}
%\end{longtable}
% \end{center} 		

\begin{table}
\caption {Log of spectroscopic observations of SN~2019wep. The phase is measured with respect to $V$-max (MJD$_{max}$=58828.5).
\label{tab:2019wep_spec_obs}}
\begin{center}
%\smallskip
%\small\addtolength{\tabcolsep}{-2pt}
\begin{tabular}{c c c c c c}
\hline \hline
Phase  &    Telescope  &     Instrument  &		Range \\
       &               &                 &             \AA   \\
\hline    
-2.9      &   2.0m LJT    &    YFOSC                  &   3400-7600           \\
-2.9      &   2.0m LJT    &    YFOSC                  &   3400-7600   \\
-2.0      &   2.0m FTN    &    FLOYDS spectrograph    &   3400-9500             \\
-1.0      &   2.0m FTN    &    FLOYDS spectrograph    &   3400-9500   \\
0.0     &   2.0m FTN    &    FLOYDS spectrograph    &   3400-9700 \\
1.4      &   2.0m LJT    &    YFOSC                  &   3400-7600  \\
5.0     &   2.0m FTN    &    FLOYDS spectrograph    &   3400-9700   \\
7.2     &   2.0m LJT    &    YFOSC                  &   3400-7600 \\
8.0     &   2.0m FTN    &    FLOYDS spectrograph    &   3400-9500\\
12.3     &   2.0m LJT    &    YFOSC                  &   3400-7600 \\
13.0     &   2.0m FTN    &    FLOYDS spectrograph    &   3400-9500 \\
18.3     &   2.0m LJT    &    YFOSC                  &   3400-7600 \\
\hline
\end{tabular}
\end{center}
\end{table}

The spectroscopic observations of SN 2019wep were taken at 9 epochs spanning up to $\sim$22 days after explosion with the 2.0m FTN and FTS telescopes of LCO and 2.4m LJT of Yunnan Observatories, China. The 1D wavelength- and flux-calibrated spectra were extracted using the \texttt{floydsspec} pipeline\footnote{\url{https://github.com/svalenti/FLOYDSpipeline}} \citep{2014MNRAS.438L.101V} for the LCO data. Spectroscopic data of the LJT telescope was reduced using the APALL task in IRAF\footnote{Image Reduction and Analysis Facility} followed by wavelength and flux calibration. The slit loss corrections were done by scaling the spectra to the photometry. Finally, the spectra were corrected for the heliocentric redshift of the host galaxy. The log of spectroscopic observations is given in Table \ref{tab:2019wep_spec_obs}. 

\section{Photometric Evolution of SN~2019\textsc{wep}}
\label{3}

\begin{figure}
	\begin{center}
		\includegraphics[width=\columnwidth]{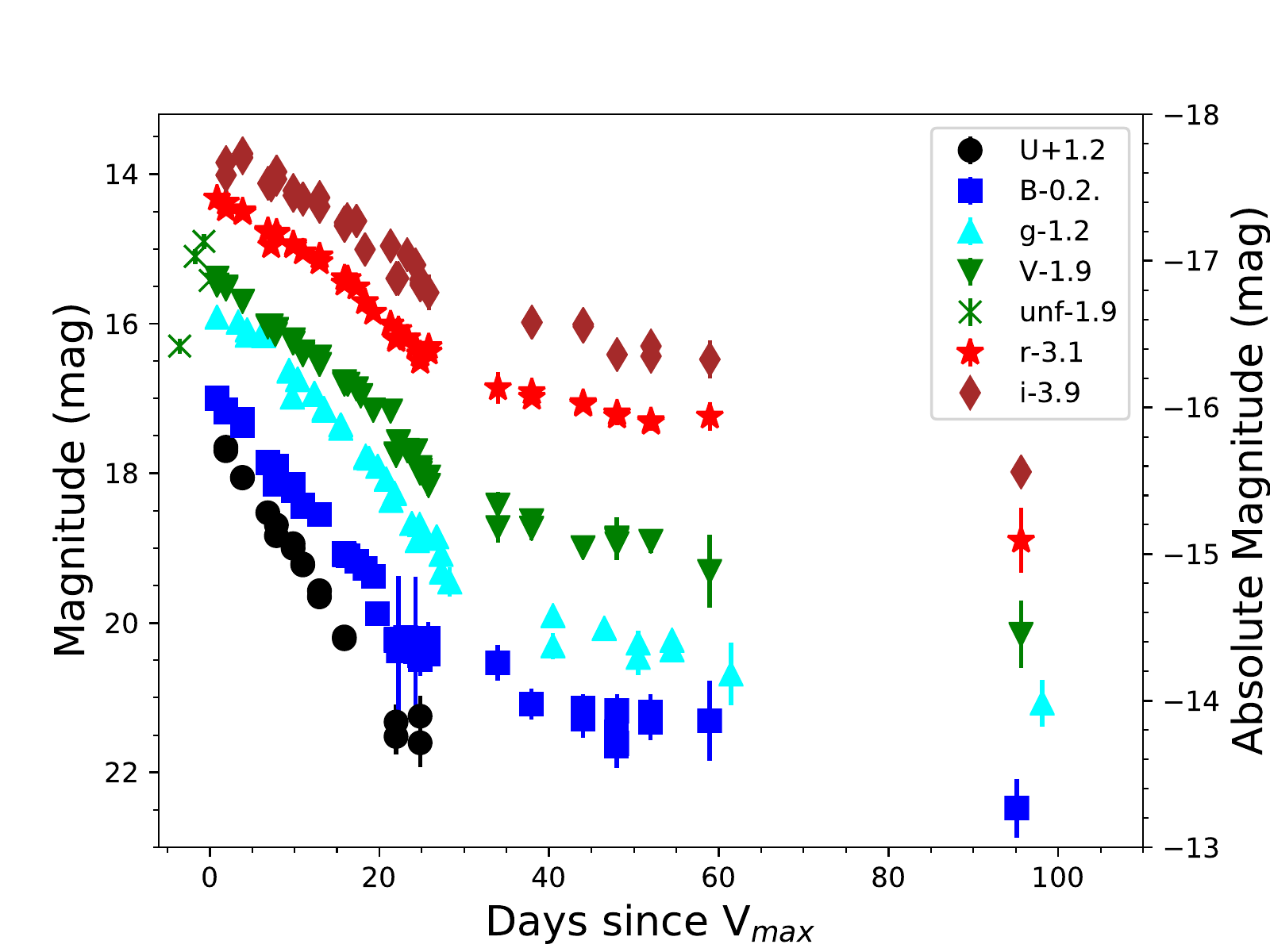}
	\end{center}
	\caption{{\it UBgVri/unfiltered} light curve evolution of SN~2019wep.}
	\label{fig:lc_2019wep}
\end{figure}

The multi-band light curve evolution of SN~2019wep is shown in Figure~\ref{fig:lc_2019wep}. Even though our observations were unable to trace the light curve maximum, four pre-maximum unfiltered data points were reported in TNS which we combine with our $V$-band data to perform a polynomial fit on the early data upto 8 days. 
A derivative of the flux values tending to zero is used to estimate the light curve maximum which occurs at MJD 58828.5 $\pm$ 2 days. The error in the obtained values is the fitting error associated with the light curve. The maximum magnitude in $V$-band is estimated by fitting a cubic spline function in the light curve data upto $\sim$  10 days.  This results in a peak absolute magnitude of M$_{V}$= -18.26 $\pm$ 0.20 mag where the estimated error accounts for the contribution of error in magnitude, redshift and interpolation errors. Even though most of the SNe~Ibn have covered the $r$-band maximum, which is also the case of our comparison sample, we use days since $V$-maximum as a reference epoch throughout the study because of a better sampling in $V$-band. 

For estimating the peak absolute magnitude in $r$- and $i$-bands, we select the first 30 days phase bin and perform a linear fit on the light curve as suggested by \cite{2017ApJ...836..158H} where the intercept gives a measure of the peak absolute magnitude. The peak absolute magnitudes for SN 2019wep are estimated to be M$_{r}$= -18.18 $\pm$ 0.95 mag and M$_{i}$= -17.98 $\pm$ 0.99 mag. The error on the peak magnitude is estimated from the estimated error on the peak date, the standard errors on the slope and intercept, and the slope-intercept covariance of the fit. We want to remark that the extrapolated peak only provides a limit on the magnitudes, and thus large error bars are associated with the estimation. 

\begin{figure*}
	\begin{center}
	    \includegraphics[width=\textwidth]{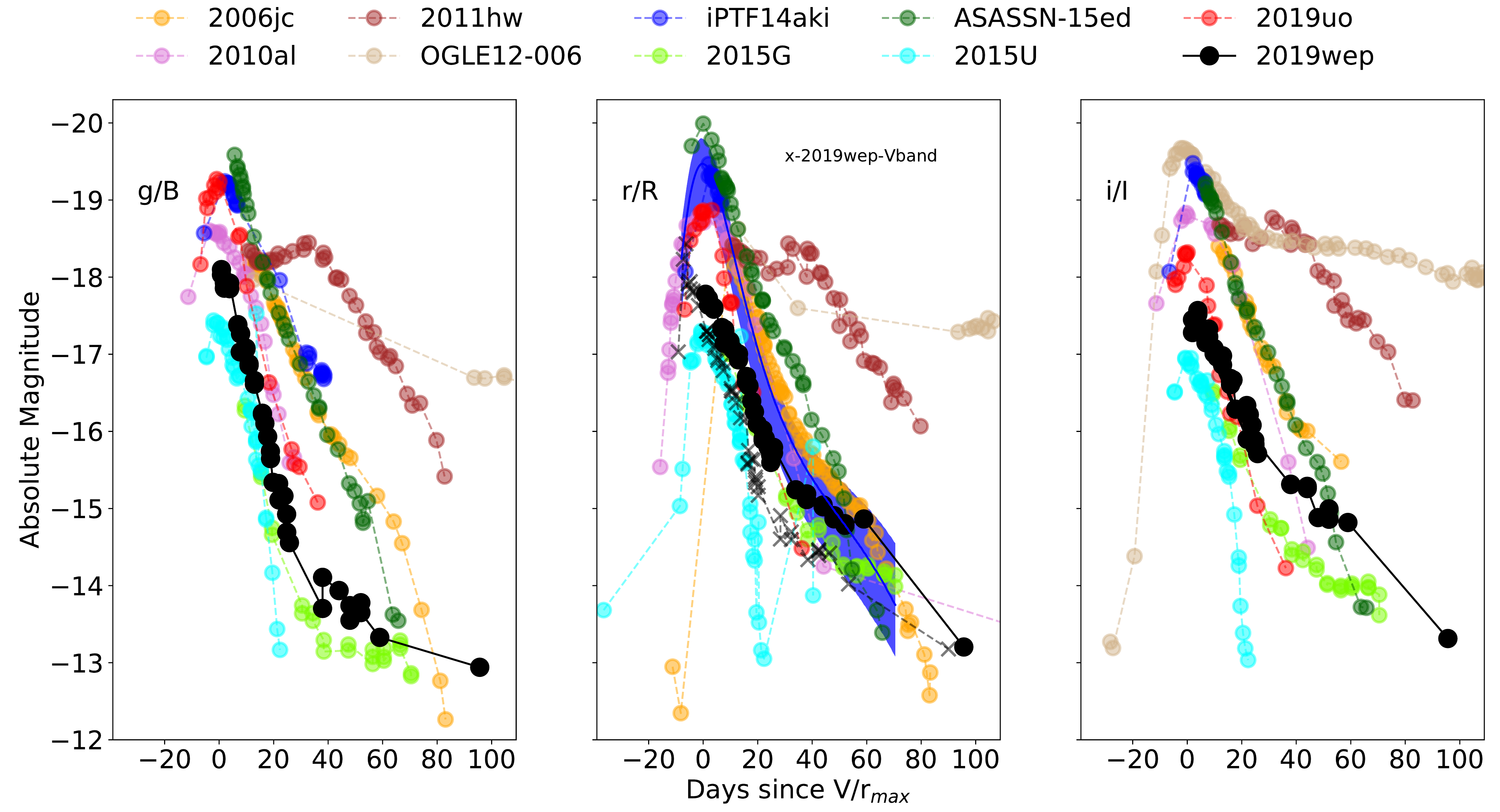}
		\includegraphics[width=\textwidth]{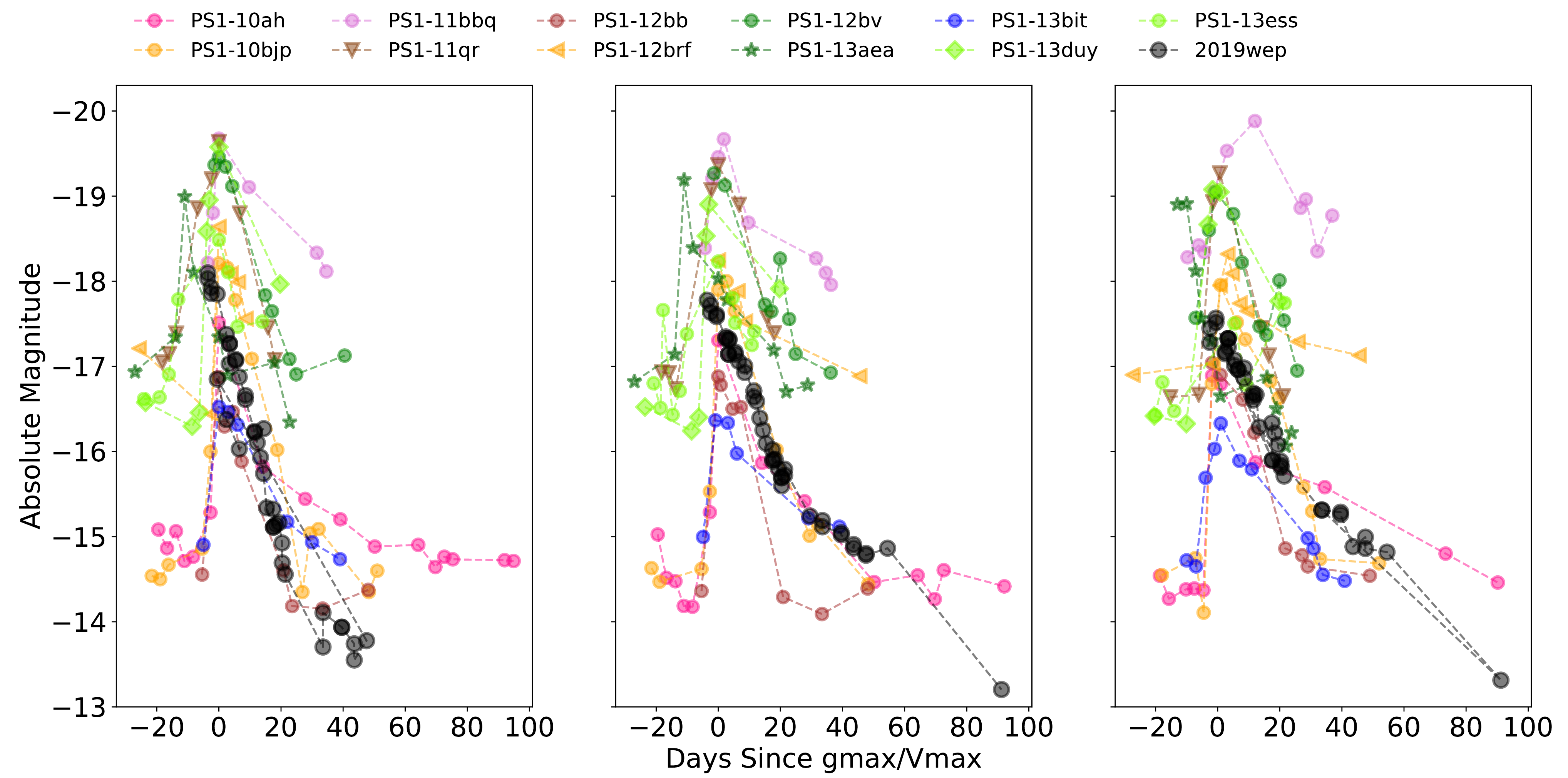}
	\end{center}
	\caption{{\it gri} light curve comparisons of SN 2019wep with a sample of SNe~Ibn \citep{2017ApJ...836..158H} on the top three panels and with a group of fast transients in the bottom three panels taken from \cite{2014ApJ...794...23D}. The SNe~Ibn comparison sample includes SNe~2006jc \citep{2007Natur.447..829P,2007ApJ...657L.105F}, 2010al \citep{2015MNRAS.449.1954P}, OGLE-SN-006 \citep{2015MNRAS.449.1941P}, 2011hw \citep{2015MNRAS.449.1954P}, iPTF14aki \citep{2017ApJ...836..158H}, 2015U \citep{2016MNRAS.461.3057S,2017ApJ...836..158H}, 2015G \citep{2017ApJ...836..158H}, ASASSN-15ed \citep{2015MNRAS.453.3649P} and 2019uo \citep{2020ApJ...889..170G}. Phase is plotted with respect to maxima for all the panels.}
	\label{fig:fasttransients}
\end{figure*}

The rise time of an object is generally estimated as the difference between the estimated explosion epoch and the maximum. In the case of SN~2019wep we estimate the rise time to be 4 $\pm$ 3 days.  The error in rise time is the error in the explosion epoch and the epoch of maximum propagated in quadrature.
The rise time of SN~2019wep indicates that it is one of the fast-rising members among the SNe~Ibn subclass and has a good match with SN~1999cq (cf. \ Table 4; \citealp{2017ApJ...836..158H, 2000AJ....119.2303M}).

The $r$-band light curve of SN~2019wep, between 0--30~days post maximum, has a decay rate of $0.088 \pm 0.002$~mag~d$^{-1}$. The $V$ and $i$-band light curves follow approximately the same decline rate as $r$-band and roughly comparable to the SNe~Ibn sample, i.e., 0.1~mag~d$^{-1}$, of course with few exceptions \citep{2017ApJ...836..158H}. The $g$- and $B$-band light curves decay slightly faster with a rate of $\sim$ 0.145 $\pm$ 0.003~mag~d$^{-1}$. Given the peak luminosity and decline rate, SN 2019wep also seems to follow the peak luminosity-decline rate relation of SNe~Ibn proposed by \cite{2021ApJ...917...97W} concluding that more luminous SNe tend to have slower post-peak decline rates because of different CSM configurations. 

Fast transients are the group of objects that rise and fade in brightness on timescales much shorter than those of other SNe \citep{2014ApJ...794...23D, 2016ApJ...819...35A, 2021arXiv211015370P, 2022arXiv220100955M}. Their progenitor systems are uncertain and they have a featureless blue continuum similar to most SNe~Ibn. The fast transients were an underrepresented class with a very few known examples \citep{2010ApJ...724.1396O, 2010ApJ...723L..98K} that were discovered by chance and displayed diverse observational properties. The recent high cadence surveys \cite{2018MNRAS.481..894P, 2020ApJ...894...27T, 2021ApJ...918...63A} have allowed for a systematic search of fast transients. Using the ZTF sample, \cite{2021arXiv210508811H} distinguished fast transients into three categories -  i) sub-luminous SN~IIb/Ic population, ii) luminous SN~Ibn population, and iii) fast and blue optical transients (FBOTs). These studies have shown similarities between SNe~Ibn and fast transients that are mostly powered by ejecta interacting with He-rich CSM. Moreover, SNe~Ibn and fast transients show similar rise times, peak luminosity and decay rates \citep{2020MNRAS.492.2208C,2021ApJ...910...42X}. Motivated with their similarity, we compare the absolute magnitude light curves of SN~2019wep in different bands with a sample of SNe~Ibn \citep{2017ApJ...836..158H} and fast transients \citep{2014ApJ...794...23D} in the top and bottom panels respectively of Figure~\ref{fig:fasttransients}. The comparison sample has well covered peaks in $g,r,i$-bands which is missed in SN~2019wep. The only peak coverage in SN~2019wep is in the $V$-band. We plot the $V$-band absolute magnitude along with $r$-band in the top middle-panel (Figure~\ref{fig:fasttransients}) to show the comparison in peak magnitudes. The limiting peak absolute magnitudes of SN~2019wep in the $r$ and $i$-bands are estimated by performing a linear fit as mentioned previously. We also estimate the 0--30 days decay rates of the sample of fast transients and find that it typically lies between 0.03-1~mag~d$^{-1}$. This indicates that the SN 2019wep light curves evolve faster than a prototypical SN~II i.e a SNe~IIP/IIL with a post plateau decay rate of 0.01 mag~d$^{-1}$  \citep{2016MNRAS.457..328L,2018ApJ...862..107B}, but has a evolution consistent with the categorised fast transients and lies at the faster decay end of SNe~Ibn. 

The top middle panel of Figure~\ref{fig:fasttransients} also shows the normalized light curve of SNe~Ibn (taken from \cite{2017ApJ...836..158H}; comprising 95\% of the SNe~Ibn data) by the blue shaded region in the $r$-band. The normalized light curve was generated by first normalizing all light curves by their estimated peak magnitudes and using a Gaussian process to estimate the average light curve. We see that the light curve shape of SN~2019wep is well matched with that of SNe~2010al, 2019uo and ASASSN-15ed and the brightness is comparable to iPTF14aki. The estimated peak magnitude in SN~2019wep is towards the lower brightness limit of the SNe~Ibn sample. Since non-detections were not included in generating the normalized light curve by \cite{2017ApJ...836..158H}, the template is biased toward a brighter and shallower evolution at early times with longer rise times of SNe~Ibn. In comparison to the fast transients, SN~2019wep lies at an average brightness. The light curve shape matches well with one of the fast transient PS1-10ah.

\begin{figure*}
\begin{center}
	    \includegraphics[scale=0.55]{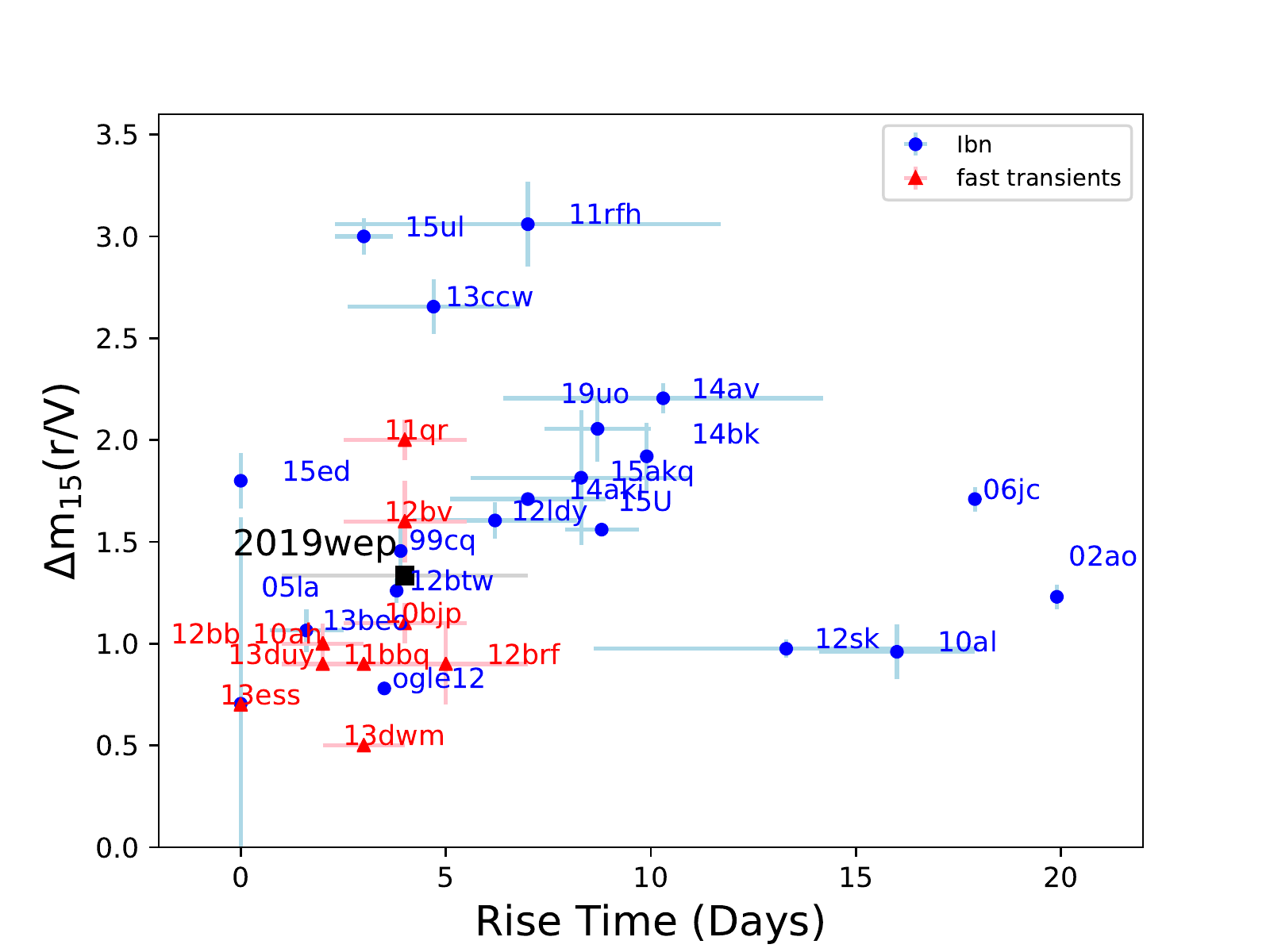} 
		\includegraphics[scale=0.55]{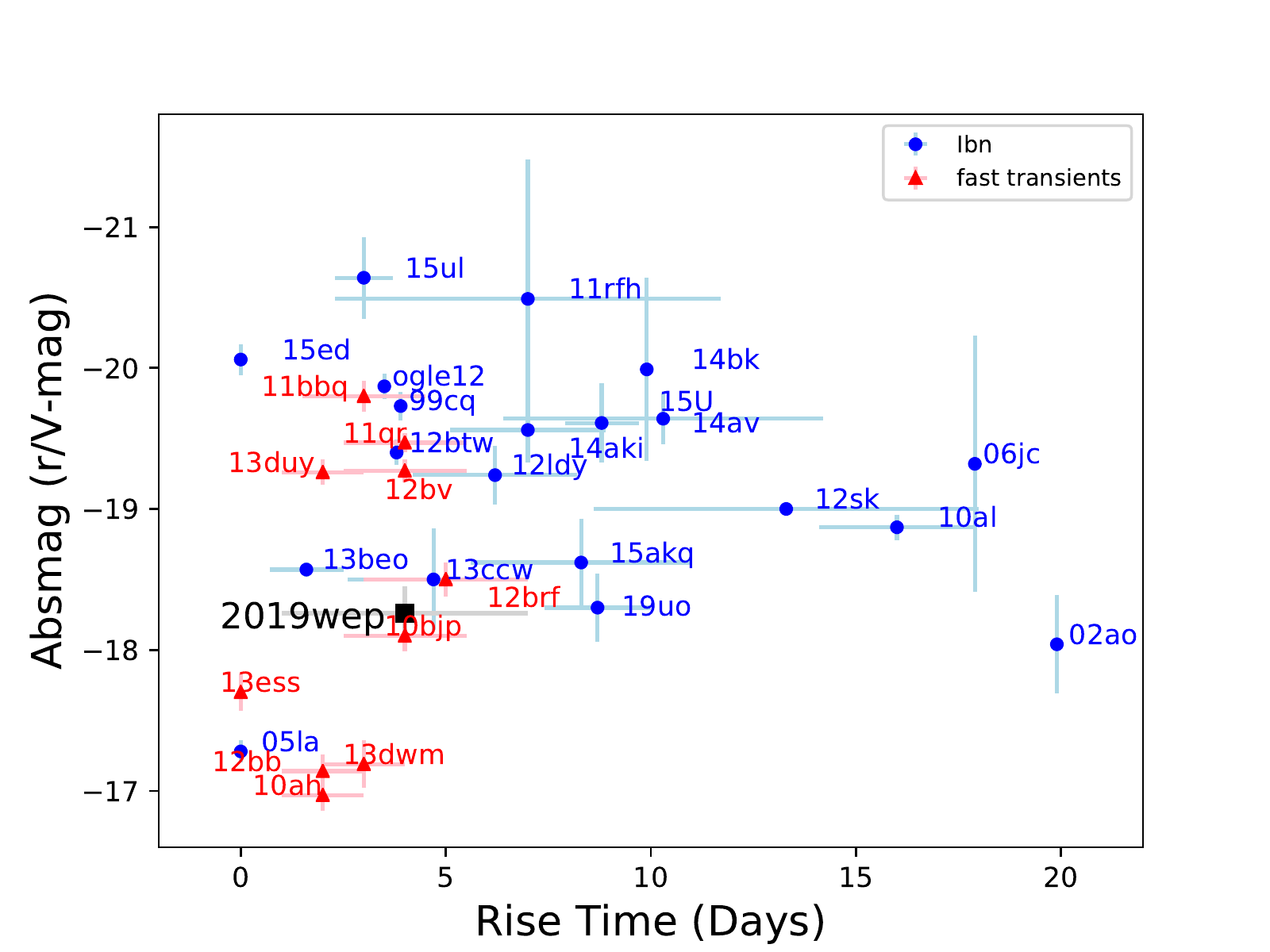}
\end{center}
	\caption{The $\Delta$m$_{15}$ vs rise time (left panel) and absolute magnitude vs rise time (right panel) correlation plots of SN 2019wep with a sample of SNe~Ibn and fast transients. The sample of type SNe~Ibn has been taken from \cite{2017ApJ...836..158H,2020ApJ...889..170G} and the sample of fast transients are taken from \cite{2014ApJ...794...23D}}
	\label{fig:correlationplots}
\end{figure*}

Figure~\ref{fig:correlationplots} shows the correlation plots of SNe~Ibn and fast transients where the $\Delta$m$_{15}$ and absolute magnitude are plotted against the rise time. The plot indicates that the rise time of SN~2019wep is shorter (even though large error bars are associated with these estimations) with respect to the rise time estimates of other SNe~Ibn. The rise time of SN~2019wep is slightly larger than the average value of the fast transients. The $\Delta$m$_{15}$ value matches well with SN~1999cq and is lower as compared to other SNe~Ibn. Even though rise times indicate that SN~2019wep lies in the faster end of the SNe~Ibn sample, still, we remark that the light curves do not evolve faster than the group of fast transients. SN~2019wep is fainter than most of the SNe~Ibn whereas it is of average brightness with respect to the fast transients. 

\begin{figure}
	\begin{center}
		\hspace{-1.0cm}
		\includegraphics[width=0.5\textwidth]{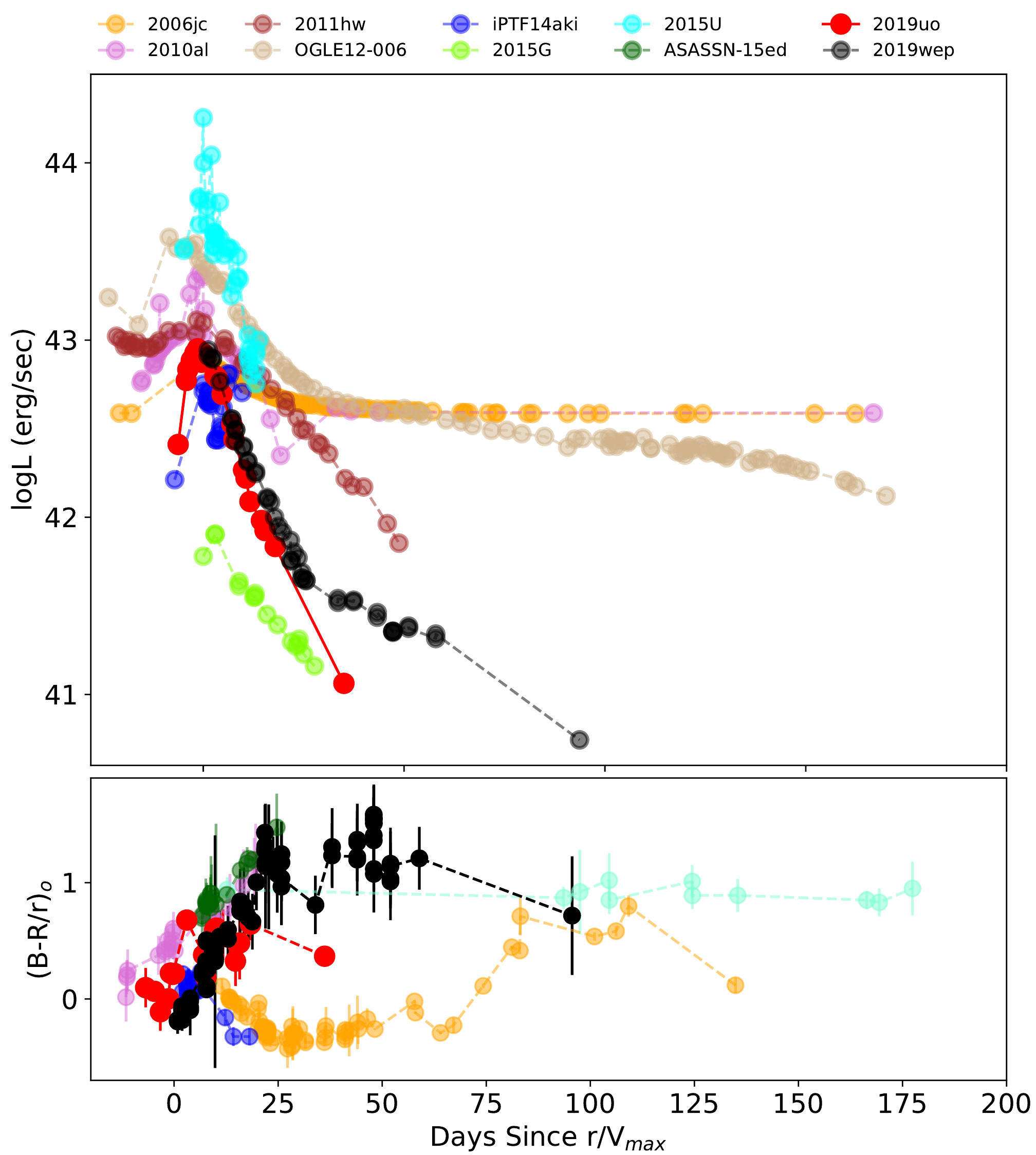}
	\end{center}
	\caption{Complete bolometric light curve and $B-R/r$ colour curve of SN~2019wep. The comparison sample includes SNe~2006jc \citep{2007Natur.447..829P,2007ApJ...657L.105F}, 2010al \citep{2015MNRAS.449.1954P}, OGLE-SN-006 \citep{2015MNRAS.449.1941P}, 2011hw \citep{2015MNRAS.449.1954P}, iPTF14aki \citep{2017ApJ...836..158H}, 2015U \citep{2016MNRAS.461.3057S,2017ApJ...836..158H}, 2015G \citep{2017ApJ...836..158H}, ASASSN-15ed \citep{2015MNRAS.453.3649P} and 2019uo \citep{2020ApJ...889..170G}.}
	\label{fig:bol_colour}
\end{figure}

We also compare the $B-R/r$ color evolution of SN~2019wep with a number of SNe~Ibn, which usually show heterogeneity in their color evolution (bottom plot; Figure~\ref{fig:bol_colour}). We see that SN~2019wep shows a color evolution similar to SNe~2010al, ASASSN-15ed and 2019uo in early times but later on transitions to blue. The $B-r$ color of SN~2019wep increases up to 1.58~mag in $\sim$52 days post maximum, and subsequently becomes blue at $\sim$93~days. SN~2019wep continues to become redder for a longer duration of time than other SNe~Ibn. For, SNe~2010al, iPTF14aki, 2019uo, ASASSN-15ed and 2019wep the $B-r$ color increases up to $\sim$1~mag, $\sim$40 days post maximum. At similar epochs, the color evolution of SN~2006jc was extremely blue ($-0.5$~mag). The transition to the redder colours for SNe~2019wep indicates that it shows the characteristic of not a typical Ibn SN, but the one which shows a close resemblance to a SN~Ib with time. This is in good agreement with the P~cygni spectroscopic features that transition from narrow to broad, indicating a He-rich circumstellar shell around the progenitor star along with optically thick CSM. The late redder colors for SN~2006jc \citep{2007Natur.447..829P} and OGLE-2012-SN-006 \citep{2015MNRAS.449.1941P} are indicative of dust formation.

To construct the bolometric light curve of SN~2019wep between UV to IR bands, we used the \texttt{SuperBol} code \citep{2018RNAAS...2..230N}. The missing UV and NIR data was supplemented by extrapolating the Specrtral Energy Distributions (SEDs) using the blackbody approximation and direct integration method as described in \cite{2017PASP..129d4202L}. A linear extrapolation was performed in the UV regime at late times. The estimated limiting peak bolometric luminosity of SN~2019wep is $8.8 \times 10^{42}$~erg~s$^{-1}$ (top panel; Figure~\ref{fig:bol_colour}). The bolometric light curves of other SNe~Ibn are also shown in the Figure. The overall light curve of SN~2019wep matches well with that of SN~2019uo (peak luminosity $\sim$ $8.9 \times 10^{42}$~erg~s$^{-1}$. \cite{2020ApJ...889..170G} inferred that SN~2019uo was best described with a $^{56}$Ni + CSM model which can be assumed to be consistent with SN~2019wep (given that both SNe have similar peak luminosities) with a $^{56}$Ni mass of $\sim$ $0.01_{-0.002}^{+0.003}$ M$_{\odot}$.  

 The bolometric light curve modelling of SN~2019wep has been performed in the work by \cite{2021arXiv211015370P} which we briefly summarise in this section. The physical parameters were derived from a best-fit hybrid model using the MINIM code \citep{2013ApJ...776..129C}, which finds self-similar solutions to the propagation of forward and reverse shocks generated by the interaction between the SN ejecta and optically thick CSM as described by \citet{1982ApJ...259L..85C}. The CSM has as its free parameters an inner radius R$_{\text{ej}}$, mass M$_{\text{CSM}}$, and density $\rho_{\text{CSM}}$, which has an assumed power law form with index $n$ = 12 in the outer region, and $n$ = 0 in the inner region. In addition to the CSM interaction, radioactive decay of $^{56}$Ni is used as an additional luminosity source. Constant optical and gamma-ray opacities are also assumed.

\begin{figure*}
	\begin{center}
		\hspace{-1.0cm}
		\includegraphics[width=0.5\textwidth]{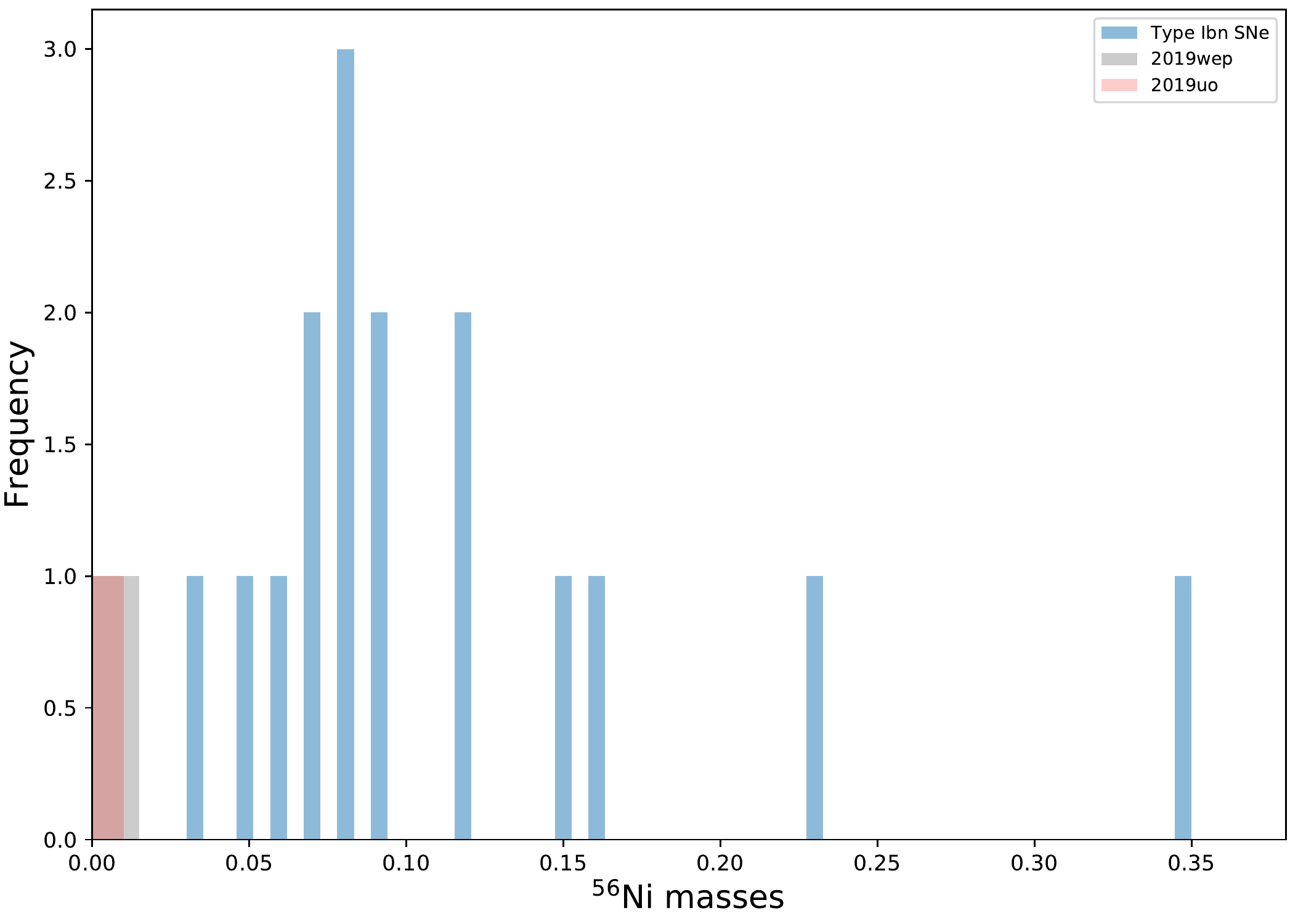}
		\includegraphics[width=0.5\textwidth]{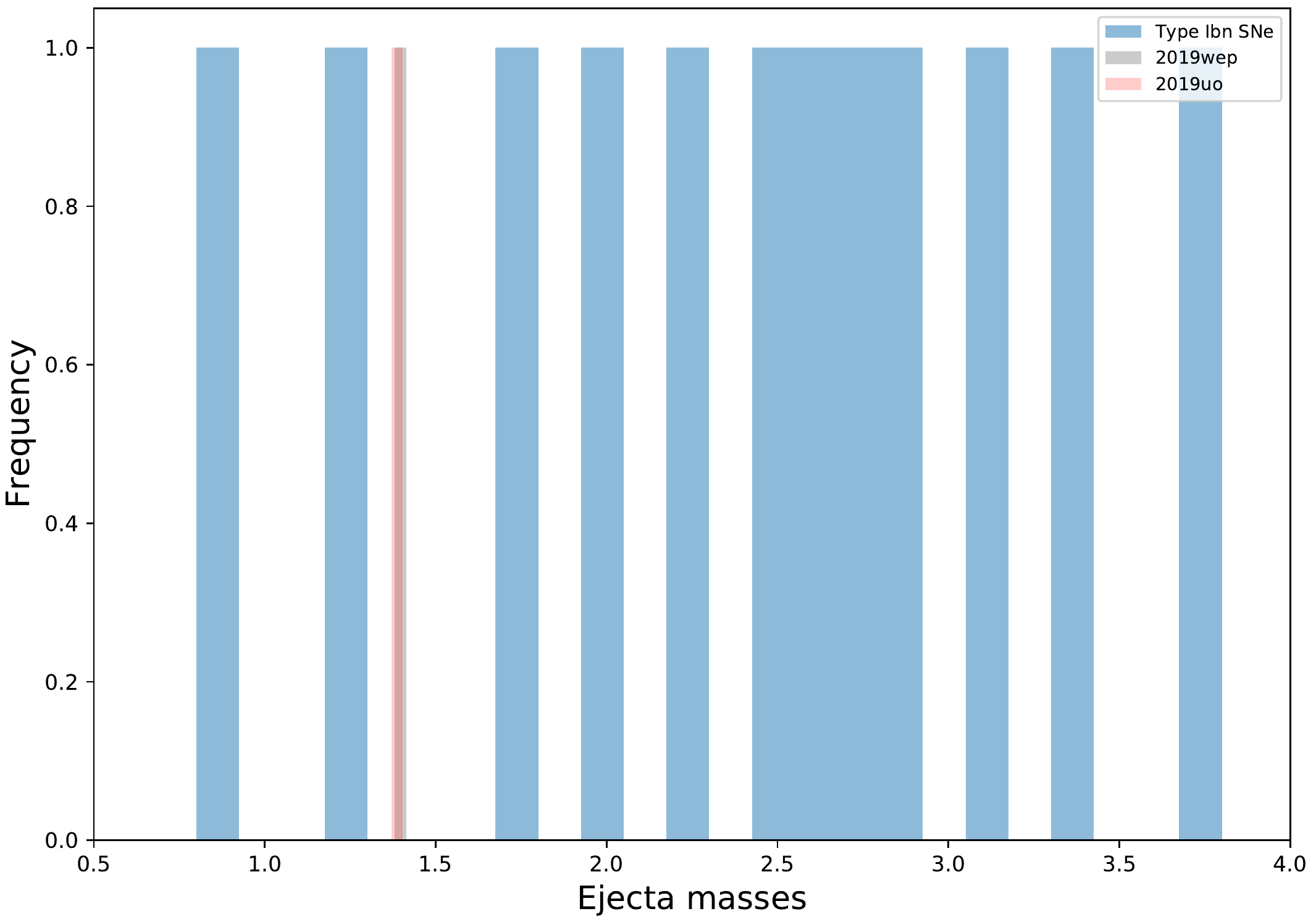}
	\end{center}
	\caption{Histograms showing the comparison of the $^{56}$Ni masses and the ejecta masses of SNe~2019wep and 2019uo with a statistical sample of SNe Ib taken from \cite{2018A&A...609A.136T,2019MNRAS.485.1559P}.}
	\label{fig:hist_Ib}
\end{figure*}

 \cite{2021arXiv211015370P} find that the bolometric evolution of SN~2019wep can be well explained by CSM interaction plus radioactive decay, following the mechanism by \citet{2013ApJ...776..129C}. The best-fit model light curve to the data is shown in their Figure 1, with model parameters listed in their Table 6 \citep{2021arXiv211015370P}. This model has very similar CSM and ejecta parameters to that of SN~2019uo, including values of R$_{\text{ej}}$, ejecta mass M$_{\text{ej}}$, M$_{\text{CSM}}$, $^{56}$Ni mass M$_{\text{Ni}}$, and $\rho_{\text{CSM}}$ that are consistent with one another. The best-fit parameters of SN~2019uo from \citet{2013ApJ...776..129C} differ from those presented in \citet{2020ApJ...889..170G} due to the fact that different assumptions were made between the models used in these two studies, including assumed CSM density power law indices. Nevertheless, the inferred parameters for these two SNe~Ibn indicate that they likely share similar CSM and explosion properties which was expected because both members belong to P~cygni class of SNe~Ibn with flash ionized features. A direct comparison of SNe~Ibn with SNe~Ib is not possible because of the difference in the luminosity powering mechanism. We see, however, that even though the obtained $^{56}$Ni masses are lower than the canonical SESNe \citep{2022arXiv220100955M}, the ejecta masses obtained are consistent with many SNe~Ib as deciphered from the statistical analysis of  \cite{2016MNRAS.457..328L,2016MNRAS.458.2973P}. This is also seen in the histograms (Figure~\ref{fig:hist_Ib}) that compares SNe~2019uo and 2019wep with the statistical sample of SNe~Ib taken from \cite{2019MNRAS.485.1559P} and \cite{2018A&A...609A.136T}. We see that both SNe~2019uo and 2019wep lie at the lower limit of $^{56}$Ni mass while the ejecta masses are quite consistent with the sample. So, we can expect the appearance of P~cygni spectroscopic features and broadening of the features with time. Also, since our obtained light curves are consistent with other SNe~Ibn, the ejecta properties are expected to be within the canonical sample of SESNe \citep{2022arXiv220100955M}. 
It is interesting, however, to note that the $^{56}$Ni masses of SNe~2019uo and 2019wep are very low. Although, there is no systematic study being done which compares the $^{56}$Ni masses with the ejecta masses for a group of SNe~Ibn, in a recent study by \cite{2022ApJ...927..180P}, very low $^{56}$Ni mass ($\leq$ 0.03 M$_{\odot}$) was estimated in a SN~Icn 2021csp using late time photometric upper limits. Analysis of three recent SNe~Icn 2021csp \citep{2022ApJ...927..180P}, 2019hgp \citep{2022Natur.601..201G} and 2021ckj \citep{2021TNSAN..71....1P} has shown that the velocities, abundance patterns and luminosities of SNe~Ibn vs SNe~Icn are parallel and share a similar WR progenitor. It is likely that the low $^{56}$Ni mass yields are due to significant fallback onto the central remnant after the explosion, which only unbound a small fraction of the $^{56}$Ni produced. This could be a possible scenario if a massive WR star undergoes only a partially successful explosion. Our estimated values of $^{56}$Ni mass from the best fit model light curves are lower than those of normal SNe~Ibc but are consistent with those of SNe~Icn (such as SN~2021csp) indicating that there was substantial fallback after the explosion (as claimed by \citealt{2022ApJ...927..180P}). This scenario also explains a reduced ejecta mass, which could explain why the ejecta mass of SN~2019wep is comparable to SNe Ib even if the progenitor is massive ($\geq$ 20 M$_{\odot}$). The other scenario could be the explosion of an ultra-stripped low-mass star, but the total ejecta mass of SN~2019wep is too high to justify this case.

\section{Spectral evolution}
\label{4}
\begin{figure}
	\begin{center}
		%\vspace{-1.0cm}
		\includegraphics[width=\columnwidth]{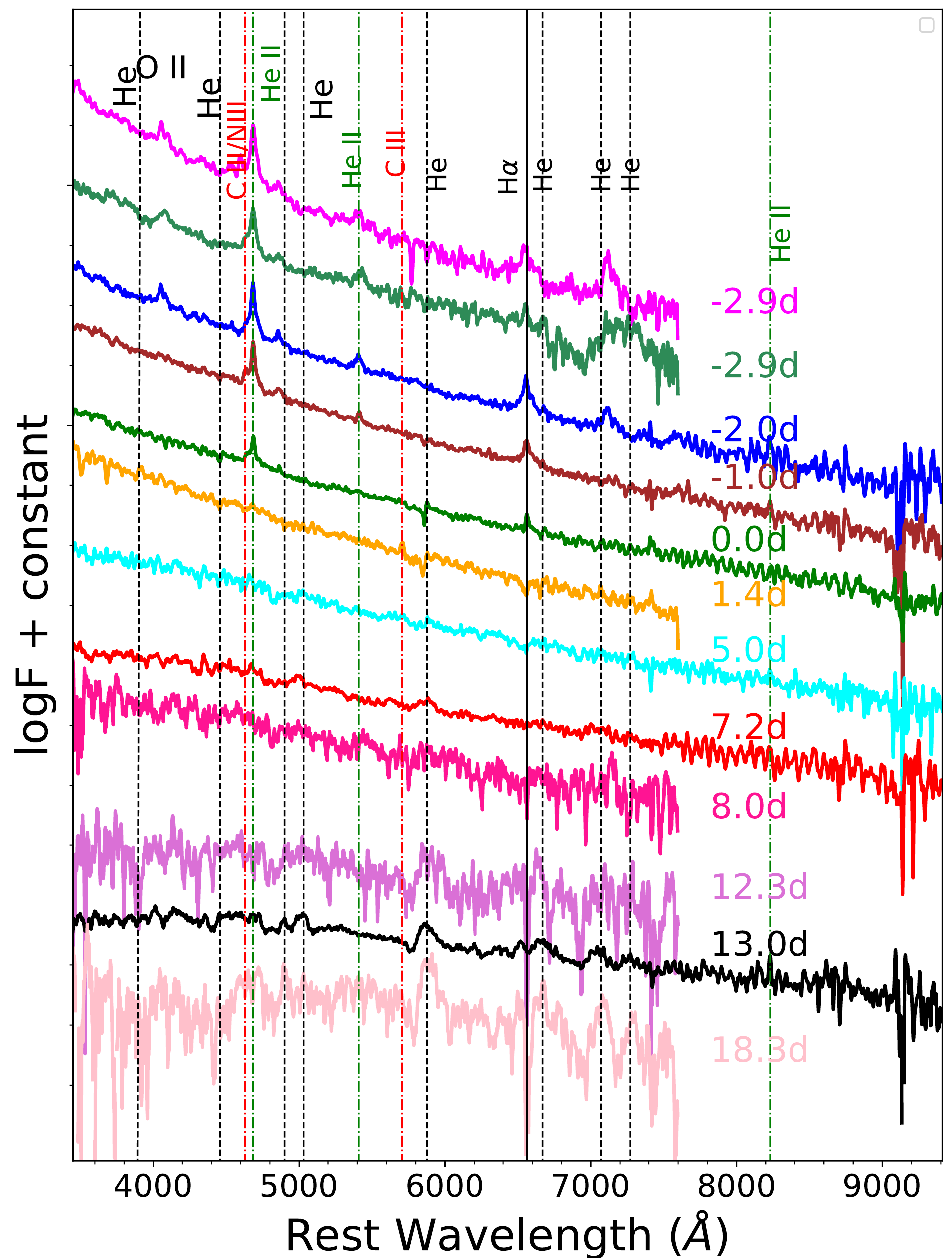}
	\end{center}
	\caption{Spectral evolution of SN~2019wep from $-2.9$~days to $18.3$~days post maximum. Prominent He features are seen in the early spectra. Flash ionization signatures of \ion{He}{2}, \ion{C}{3} and \ion{N}{3} are also seen. Broadening of He lines become prominent with time.}
	\label{fig:spec_early}
\end{figure}

\begin{figure}
	\begin{center}
		%\vspace{-1.0cm}
		\includegraphics[width=\columnwidth]{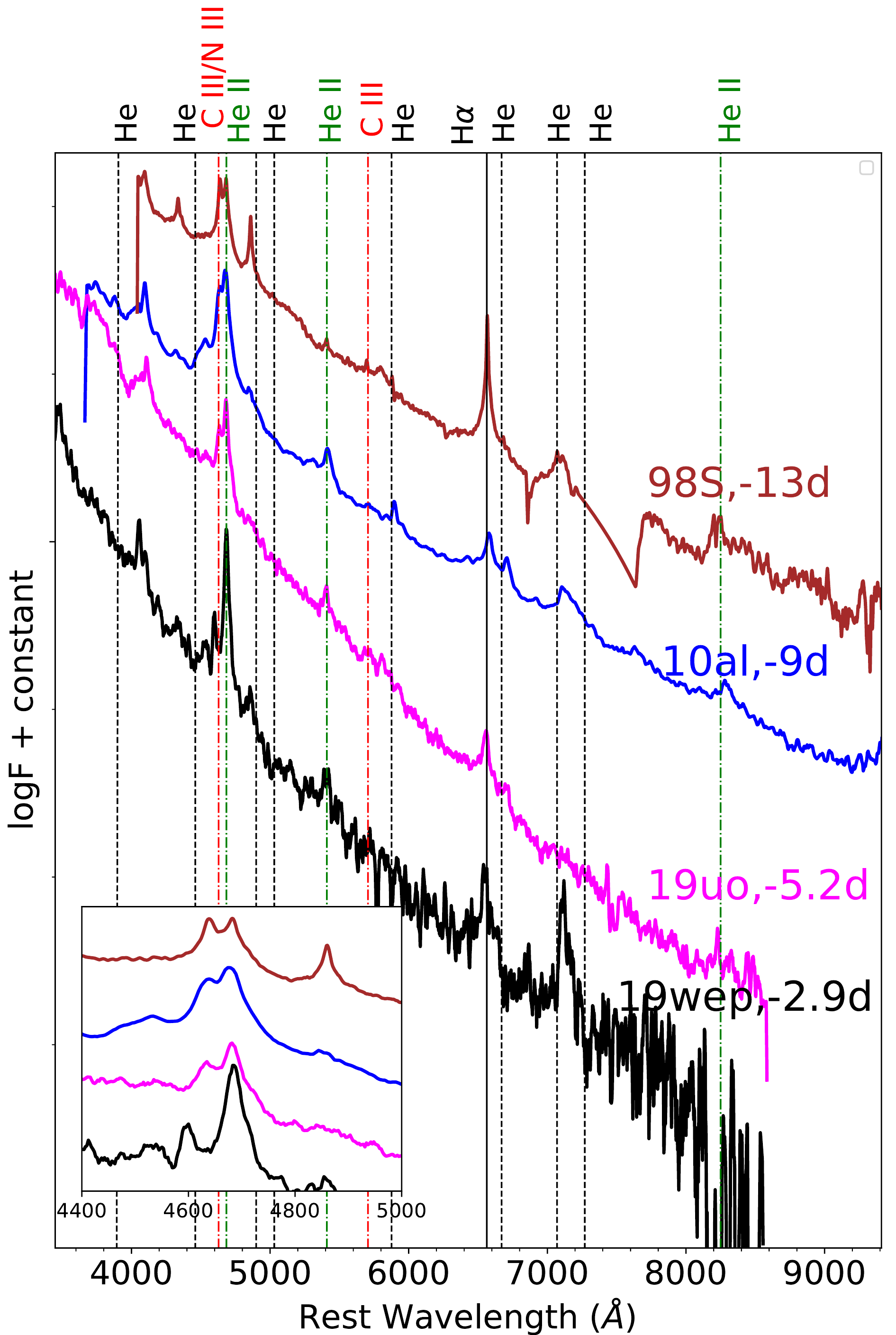}
	\end{center}
	\caption{Spectral comparison of SN~2019wep at $-2.9$~days with SNe~1998S \citep{2001MNRAS.325..907F} 2010al \citep{2015MNRAS.449.1954P} and 2019uo \citep{2020ApJ...889..170G}. Prominent flash ionization features are marked.}
	\label{fig:comp_early}
\end{figure}

We present the spectral evolution of SN 2019wep from $-2.9$~days to $18.3$~days post maximum (Figure~\ref{fig:spec_early}). The first few spectra of SN~2019wep are very similar to SNe~2010al and 2019uo with a distinct blue continuum. The presence of this blue continuum also makes it similar to fast transients (\citealt{2014ApJ...794...23D,2021arXiv211015370P}).We fit a blackbody on the first three spectra ($-2.9$, $-2.0$ and $-1.0$ day) of SN 2019wep which show that the photospheric temperature varies between 22,000 K to 15,000 K. However, we want to remark that the SEDs of these SNe peak in the UV regime \citep{2014ApJ...794...23D} and the temperature will essentially depend on the UV brightness of the object. As such, the temperature derived from the photometry is uncertain owing to the lack of UV observations. A narrow H$\alpha$ emission is seen in the early spectrum of SN~2019wep with a velocity of $\sim$ 3000 km sec$^{-1}$ which is most likely due to interstellar gas in the host galaxy and some contribution from SN which we will explain in detail in later sections. We see typical features of emission in SN~2019wep around $\sim$4660 \AA~ in the first five spectra ($-2.9$ to $0.0$~days). The emission components are double-peaked with two distinct transitions, the blue component peaking at 4643 \AA\ and the red component peaking at 4682 \AA. The red component at 4682 \AA\ is due to \ion{He}{2} at 4686 \AA, whereas the blue component arises from a blend of \ion{C}{3} 4648 \AA\ and \ion{N}{3} 4640 \AA. A very weak, less prominent doubly ionized \ion{C}{3} feature at 5696 \AA\ is also seen. These signatures are interpreted as He rich signatures in a flash ionised CSM \citep{2014AAS...22323502G,2015MNRAS.449.1921P}. Previously flash ionisation signatures of \ion{C}{3} were seen in PTF12ldy and iPTF15ul \citep{2017ApJ...836..158H}. In SNe~Ibn 2010al and 2019uo the flash ionization signatures of both \ion{C}{3} and \ion{He}{2}, typical of SNe~II, are seen. Presence of such lines indicate that they could likely be originated from WR winds \citep{2010ATel.2491....1C, 2010CBET.2223....1S}. Flash ionized signatures were previously noted in SNe~IIn (e.g., SN~1998S; \citet{2001MNRAS.325..907F} and SN~2008fq; \citet{2013A&A...555A..10T}). \cite{2014A&A...572L..11G} reverted the WR progenitor scenario through the estimation of the physical parameters of a SN~IIb 2013cu and found that the progenitor is not a WR star even though the flash ionized features showed WR like lines. A \ion{He}{2} 5411 \AA\ feature with a velocity between  1413~km~s$^{-1}$--1483~km~s$^{-1}$ is also seen in the very early spectra ($-2.9$-$0.0$ day) of SN~2019wep. In the early spectra, an absorption feature is seen close to $\sim$4000 \AA\ which is most likely due to the presence of \ion{O}{2} and \ion{He}{2} features, respectively.

Figure~\ref{fig:comp_early} shows a comparison of the early spectra of SNe~1998S (type~IIn), 2010al (type~Ibn) and 2019uo (type~Ibn) with SN~2019wep. It is interesting to note that SN~2019wep shows the strongest flash features in the early spectrum as compared with other SNe~Ibn. The spectrum of SN~2010al shows only \ion{C}{3} features around 4650 \AA, SN~2019wep shows \ion{C}{3} features around 4650 \AA\ and at 5696 \AA\ in agreement with SN~2019uo. The inset in Figure~\ref{fig:comp_early} highlights these features, which also indicates that SN~2019wep shows the most significant \ion{He}{2} feature.

\begin{figure}
	\begin{center}
		%\vspace{-1.0cm}
		\includegraphics[width=\columnwidth]{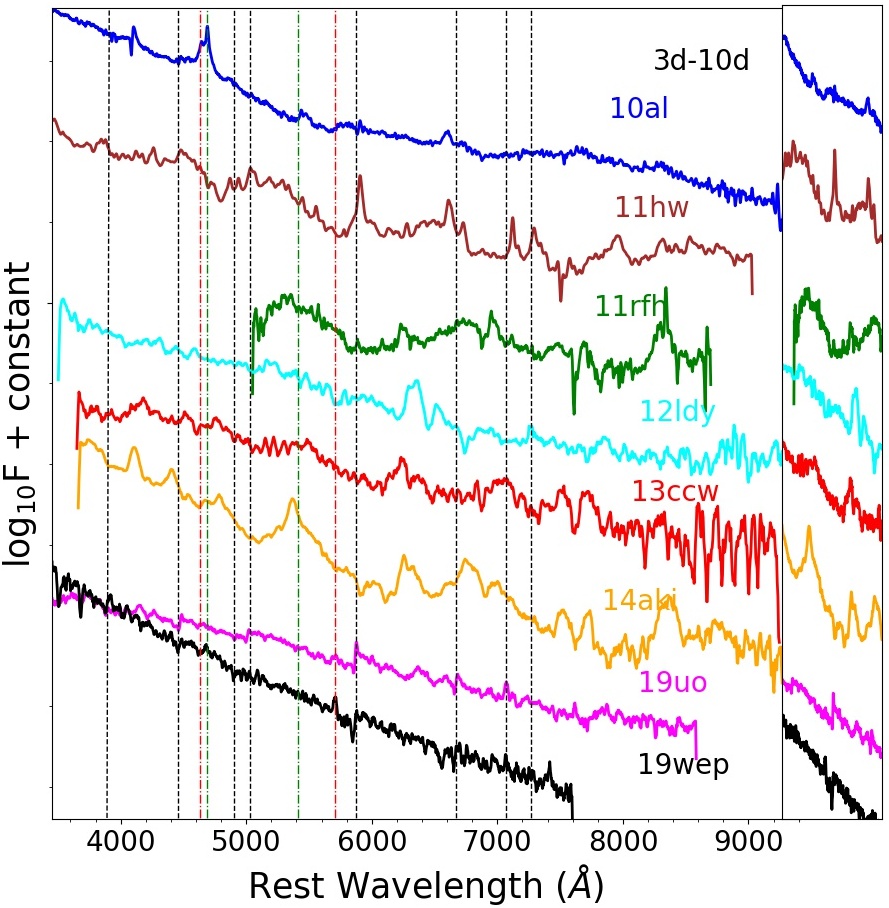}
	\end{center}
	\caption{Comparison of the spectrum of SN~2019wep to other SNe~Ibn. SN~2019wep belongs to the P~cygni subclass and shows a close resemblance to SNe~2010al and 2019uo. The data for this are taken from --- SNe~2010al \citep{2015MNRAS.449.1921P}, 2011hw \citep{2015MNRAS.449.1921P}, PTF11rfh \citep{2017ApJ...836..158H},  PTF12ldy \citep{2017ApJ...836..158H}, LSQ13ccw \citep{2015MNRAS.449.1954P}, iPTF14aki \citep{2017ApJ...836..158H} and 2019uo \citep{2020ApJ...889..170G}. The right panel of the plot shows the P~cygni and the emission lines of \ion{He}{1} 5876 \AA~ feature.}
	\label{fig:comp_mid}
\end{figure}

\begin{figure}
	\begin{center}
		%\vspace{-1.0cm}
		\includegraphics[width=\columnwidth]{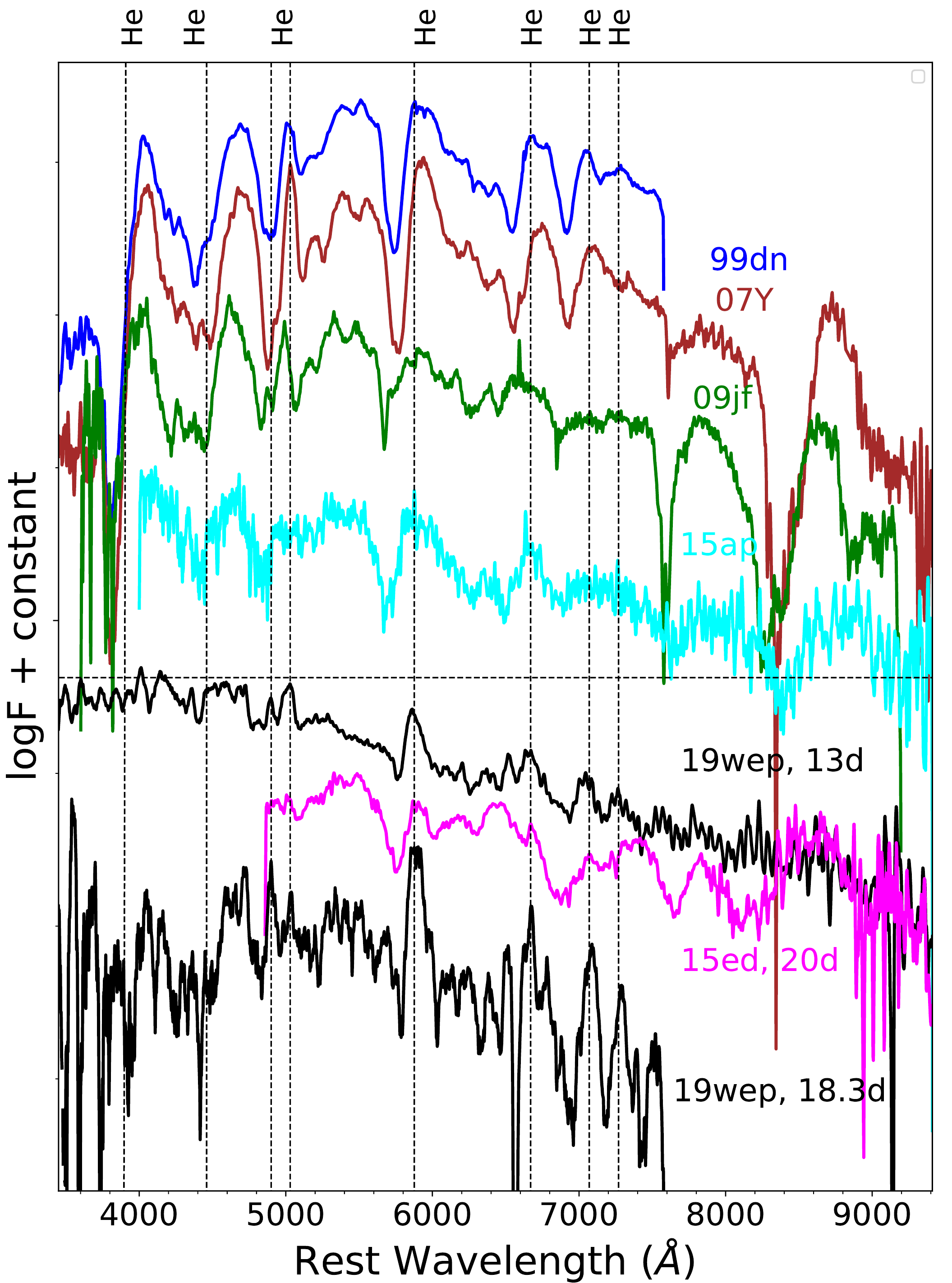}
	\end{center}
	\caption{Comparison of the spectrum of SN~2019wep to other SN~Ib and a transitioning SN~Ibn ASASSN-15ed. The data are taken from - ASASSN-15ed \citep{2015MNRAS.453.3649P}, SNe- 1999dn \citep{2011MNRAS.411.2726B}, 2007Y \citep{2009ApJ...696..713S}, 2009jf \citep{2011MNRAS.416.3138V} and 2015ap \citep{2020MNRAS.497.3770G}.}
	\label{fig:comp_Ib}
\end{figure}

\begin{figure*}
	\begin{center}
		%\vspace{-1.0cm}
		\includegraphics[scale=0.4]{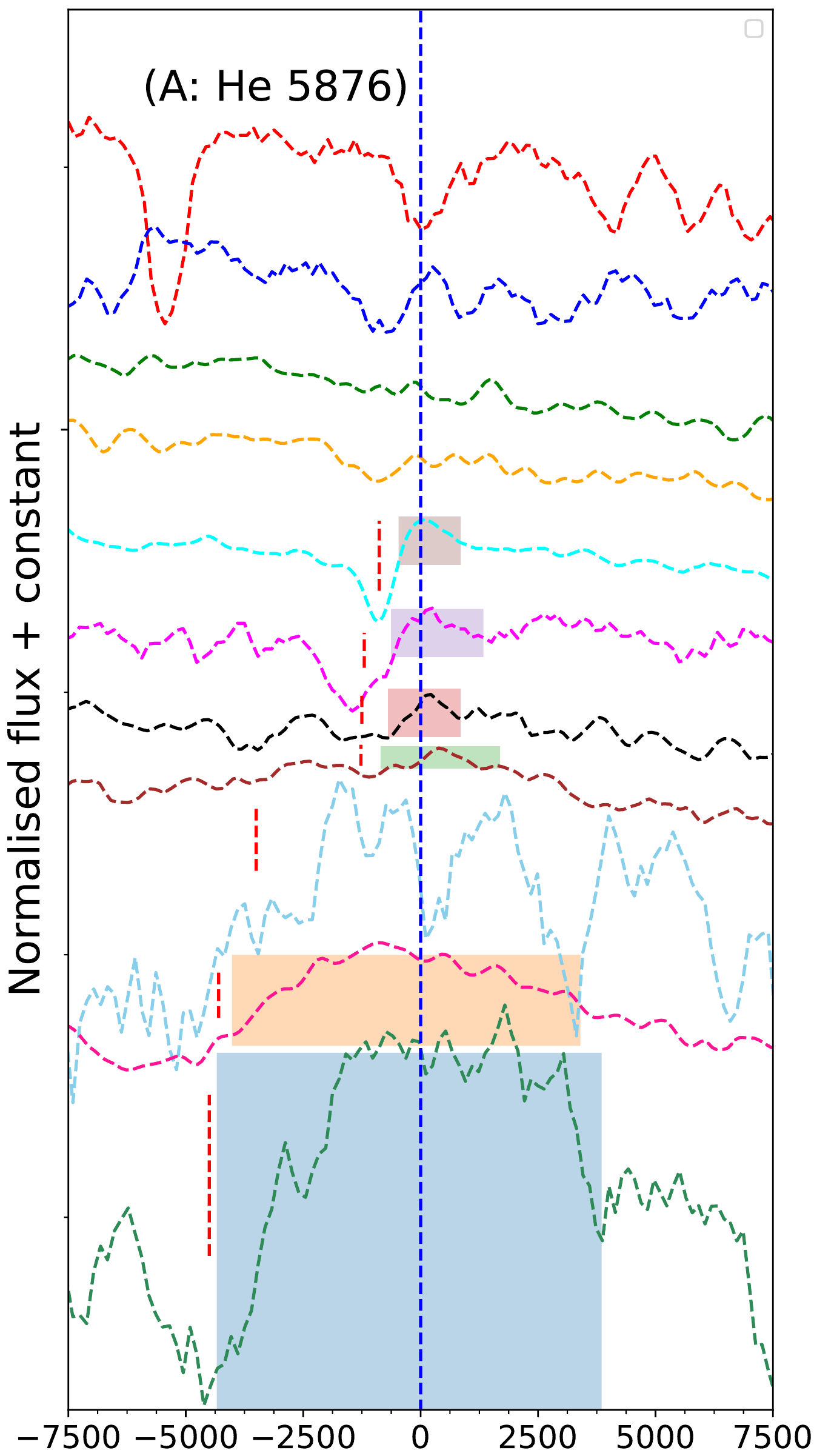}
		\includegraphics[scale=0.4]{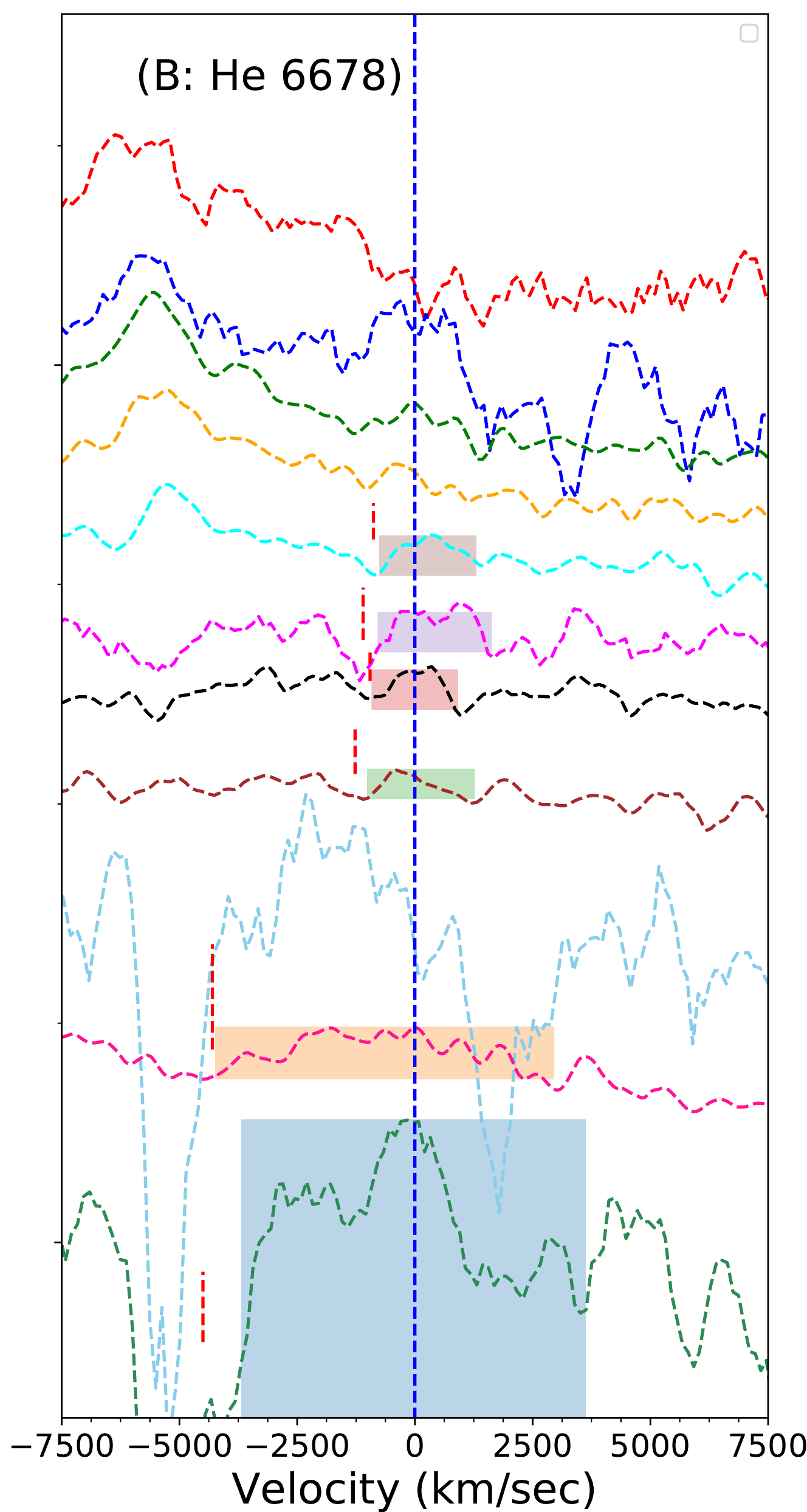}
		\includegraphics[scale=0.4]{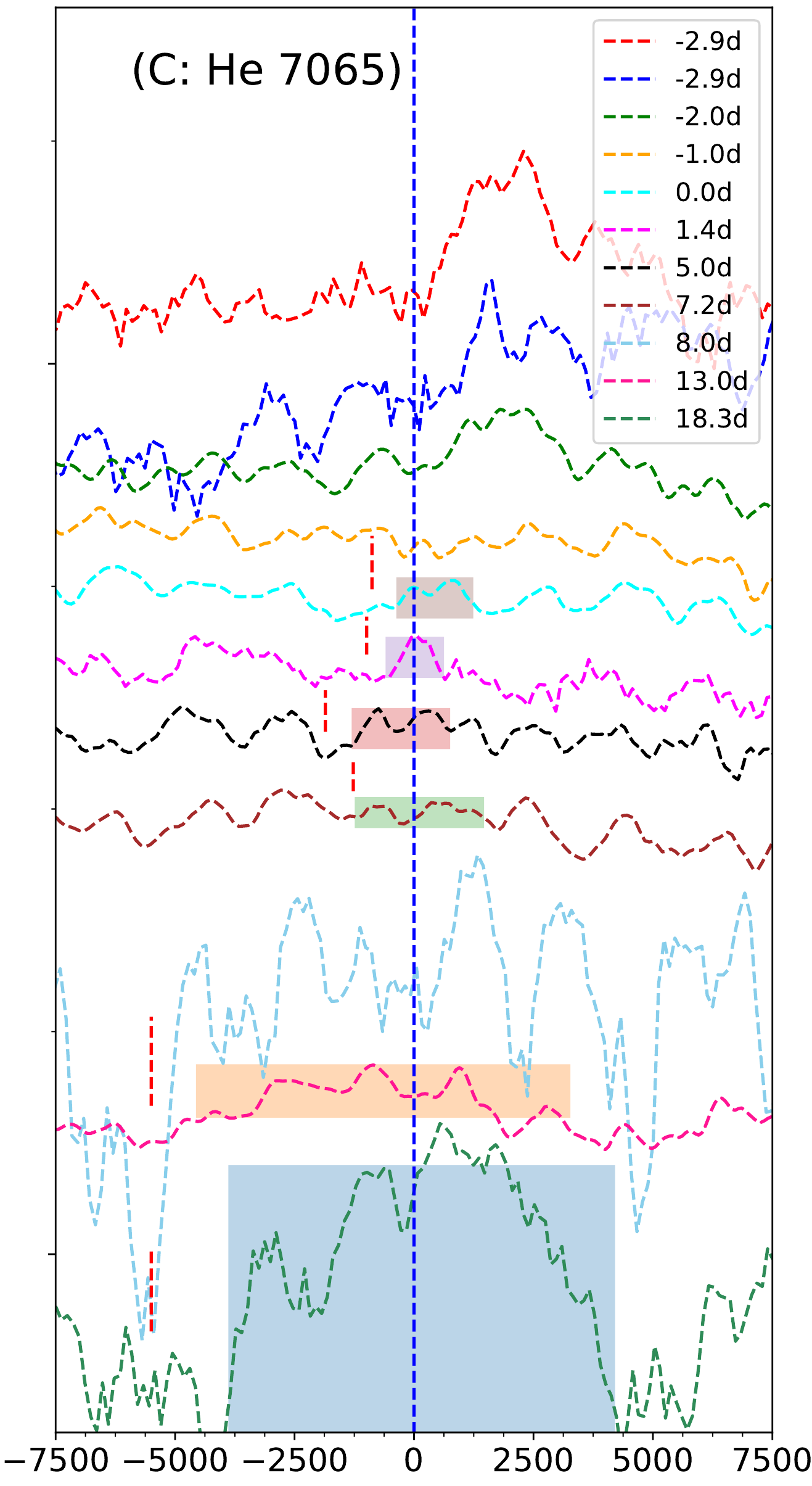}
	\end{center}
	\caption{ Line evolution of \ion{He}{1} 5876, 6678 and 7065 \AA~. These lines broaden with time showing prominent evolution. The blue dotted line corresponds to the central velocity of 0 km sec$^{-1}$. The red dotted line represents the absorption minima corresponding to the velocity estimations. The shaded regions in the plot describes the area used for the estimation of the equivalent widths.}
	\label{fig:vel_evolution}
\end{figure*}

\begin{figure*}
	\begin{center}
		%\vspace{-1.0cm}
		\includegraphics[width=1.0\textwidth]{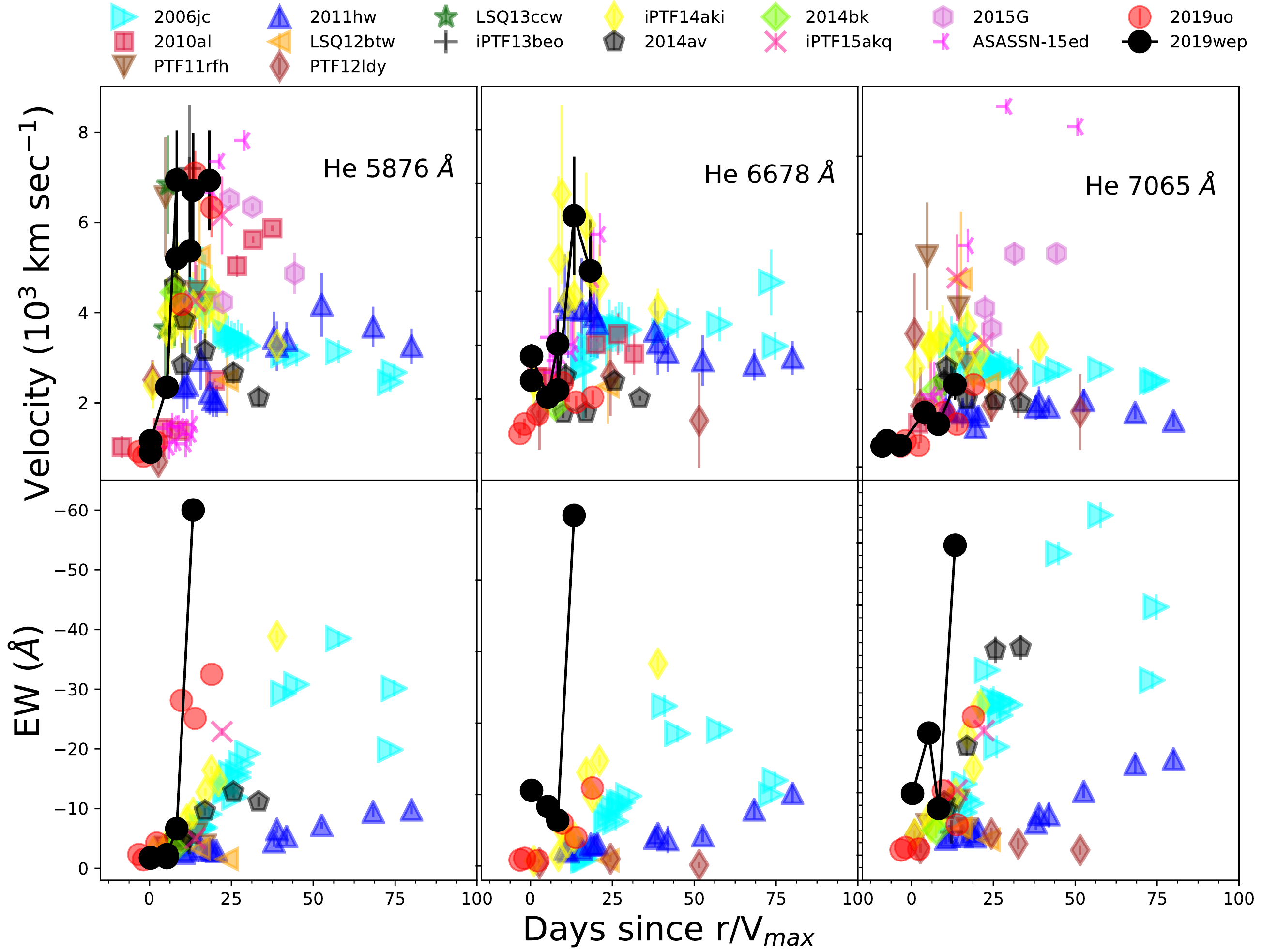}
	\end{center}
	\caption{Evolution of line velocities and equivalent widths of \ion{He}{1} emission lines is shown in top and bottom panels, respectively. The data for this are taken from --- SNe~2006jc \citep{2007ApJ...657L.105F,2008MNRAS.389..113P}, 2010al \citep{2015MNRAS.449.1921P}, 2011hw \citep{2015MNRAS.449.1921P}, PTF11rfh \citep{2017ApJ...836..158H}, LSQ12btw \citep{2015MNRAS.449.1954P}, OGLE12-006 \citep{2015MNRAS.449.1941P}, PTF12ldy \citep{2017ApJ...836..158H}, iPTF13beo \citep{2014MNRAS.443..671G}, LSQ13ccw \citep{2015MNRAS.449.1954P}, iPTF14aki \citep{2017ApJ...836..158H}, 2014av \citep{2016MNRAS.456..853P}, 2014bk \citep{2016MNRAS.456..853P}, iPTF15akq  \citep{2017ApJ...836..158H}, ASASSN-15ed \citep{2015MNRAS.453.3649P}, 2015G \citep{2017ApJ...836..158H} and 2019uo \citep{2020ApJ...889..170G}.}
	\label{fig:vel_eq}
\end{figure*}

As the SN evolves further ($0.0$~days), we see a distinct \ion{He}{1} P~cygni feature that is superimposed on a broader base (the continuum is not flat). The flash ionization spectral signatures vanishes with time and disappears on $1.4$ days. From $5.0$--$18.3$~days, features of \ion{Ca}{2}, \ion{Si}{2}, and \ion{Na}{1}D also start developing (see Figure~\ref{fig:spec_early}). Figure \ref{fig:comp_mid} shows the spectral comparison of SN~2019wep with other SNe~Ibn in the phase range between 3 day to 10 day post maximum. At similar phases, the \ion{He}{1} 5876 \AA\ feature of SN~2019wep is similar to that identified in SN~2010al and SN~2019uo with a narrow P~cygni feature while other members show a prominent emission profile. Also, the continuum of SN~2019wep is bluer as compared to all other SNe of the comparison sample.  However, the \ion{He}{1} P~cygni feature of SN~2019wep is broader than SNe~2010al and 2019uo, and is superimposed over a broader emission line. On the other hand, the \ion{He}{1} P~cygni profile in SN~2010al is over a flat continuum. The flash ionization signatures in SN~2019wep have vanished in this phase, but in SN~2010al these signatures are still visible. The spectral signatures of SN~2019wep (specially \ion{He}{1} as shown in zoomed version in the right inset) indicates that SN~2019wep belongs to the ``P~cygni" subclass. 

We see distinct broad emission lines of \ion{He}{1} developing from $7.2$-$18.3$ day indicating similarity to a SN~Ib. For a more evident comparison, we plot the spectrum of SN~2019wep in  Figure~\ref{fig:comp_Ib} with SN~Ibn ASASSN-15ed \citep{2015MNRAS.453.3649P} and a few prototypical SNe~Ib between $15$--$20$ days after maximum. A comparison of the \ion{He}{1} velocity of SN~2019wep shows that it has a typical absorption FWHM velocity of $\sim$ 5000 $\pm$ 1000 km sec$^{-1}$. At similar epochs, the FWHM velocity of ASASSN-15ed is 6300 $\pm$ 1000 km sec$^{-1}$ which is comparable to that of SN~2019wep. This is less than the typical SN~Ib velocity of our comparison sample which has absorption trough velocity between 7000-9000 km sec$^{-1}$. The major difference observed between SNe~Ibn and SNe~Ib spectra is that SNe~Ibn have more symmetric profiles, while in SNe~Ib the absorption dominates over the emission. Additionally, the absorption components of \ion{He}{1} in SN~2019wep are almost similar to SNe~Ib spectra suggesting a higher kinetic energy per mass unit in the SN ejecta, specially in the late phases of SNe~Ibn that develop broad lines \citep{2015MNRAS.453.3649P}. The simultaneous presence of a very narrow feature at early phases and transitioning to broad features suggests that these
features arise from different emitting regions: the broader \ion{He}{1} P~cygni features are likely a signature of the SN ejecta, while the narrow \ion{He}{1} P~cygni lines are generated in the unperturbed, He-rich CSM. At early phases, the narrow lines are formed in the photosphere located in a dense shell. This shell is either continuously photoionized by early ejecta-CSM interaction in the inner CSM regions and/or by the initial shock breakout. Once recombination occurs, the shell becomes transparent and we see the signatures of underlying SN ejecta and the spectrum is dominated by broad P~cygni lines. 

Figure \ref{fig:vel_evolution} shows the evolution of \ion{He}{1} 5876, 6676, 7065 \AA~ feature in the velocity space. We see a narrow P~cygni profile which gradually evolves with time and shows a broad emission on top. The physical explanation behind the origin of the ``P~cygni" subclass could be a shell of He around the progenitor star surrounded by a dense CSM. As the optically thick shell is lit by the explosion, the narrow P~cygni features transition to broader emission as the shell is swept up by the SN ejecta. The viewing angle dependence could also affect this scenario; if the CSM is asymmetric and, we have a He~rich torus, then P~cygni features would only be visible if the system is viewed edge-on, while emission features can be seen only if it is viewed face-on. On the contrary, \cite{2019arXiv191006016K} suggest that dominance of emission at late phases is not because of being optically thin, but because they lack other lines to branch into it. UV and X-ray emission arising at the shock boundary are the source of He ionisation and recombination. The ionised region outside the shock leads to electron scattering and emission, P~cygni features usually originate from optical depths $\leq$1. X-rays penetrating further into the P~cygni producing regions will fill in the absorption and lead to emission features. Thus, this provides an alternative scenario to the transitioning of P~cygni to emission features of \ion{He}{1} lines for SNe~Ibn. Asymmetric CSM, thus, plays an important role in the origin of these unusual features. Much of the ejecta usually moves undisturbed by the prevalent CSM configuration. Luminosity generation as a consequence of the interaction of the equatorial CSM beneath the photosphere continues, but, spectral signatures remain hidden until the photosphere recedes. This is the primary cause driving the blue continuum at early stages followed by the appearance of broadened redshifted He features \citep{2018MNRAS.477...74A}. 

We measure the expansion velocities and equivalent widths (EWs) of three neutral He lines (5876, 6678, and 7065 \AA), wherever visible. The emission lines of \ion{He}{1} were fit using a Gaussian on a linear continuum. The estimation of EW involves calculation of the integral of the flux normalized to the local continuum. We do not measure the EW of the P~cygni lines. The velocities reported are estimated from the absorption minima of P~cygni profiles. Figure~\ref{fig:vel_eq} shows the evolution of velocity and EW for a sample of SNe~Ibn taken from \cite{2017ApJ...836..158H}. We see an increasing trend in both the line velocities and EW of the He lines. The velocity estimates of SN~2019wep lie at the upper range of SNe~Ibn and show a faster evolution. In SN~2019wep, the broad features seen are more prominent than typical emission velocity of SNe~Ibn around maxima varying between 6000-8000 km sec$^{-1}$ \citep{2017ApJ...836..158H}. SN~2019wep has a good resemblance with ASASSN-15ed reaching broader emission profiles as seen in the P~cygni subclass \citep{2017ApJ...836..158H} while the emission subclass shows less evolution in line velocities. The EWs of the absorption component of \ion{He}{1} lines for SN~2019wep are at an extreme high end than the normal SNe~Ibn. This is also an indication that perhaps SN~2019wep lies somewhere between SN~Ib and SNe~Ibn.\\

\noindent{}
{\bf \ion{He}{1} line velocity evolution :} \\
A wide diversity and heterogeneity is observed in the spectral evolution of SNe~Ibn in accordance with SNe~IIn. The physical parameters of SNe~Ibn strongly depend on the composition, geometry, mass and density profile of CSM along with the residual stellar envelope at the time of explosion.

\begin{figure}
	\begin{center}
		%\vspace{-1.0cm}
		\includegraphics[scale=0.35]{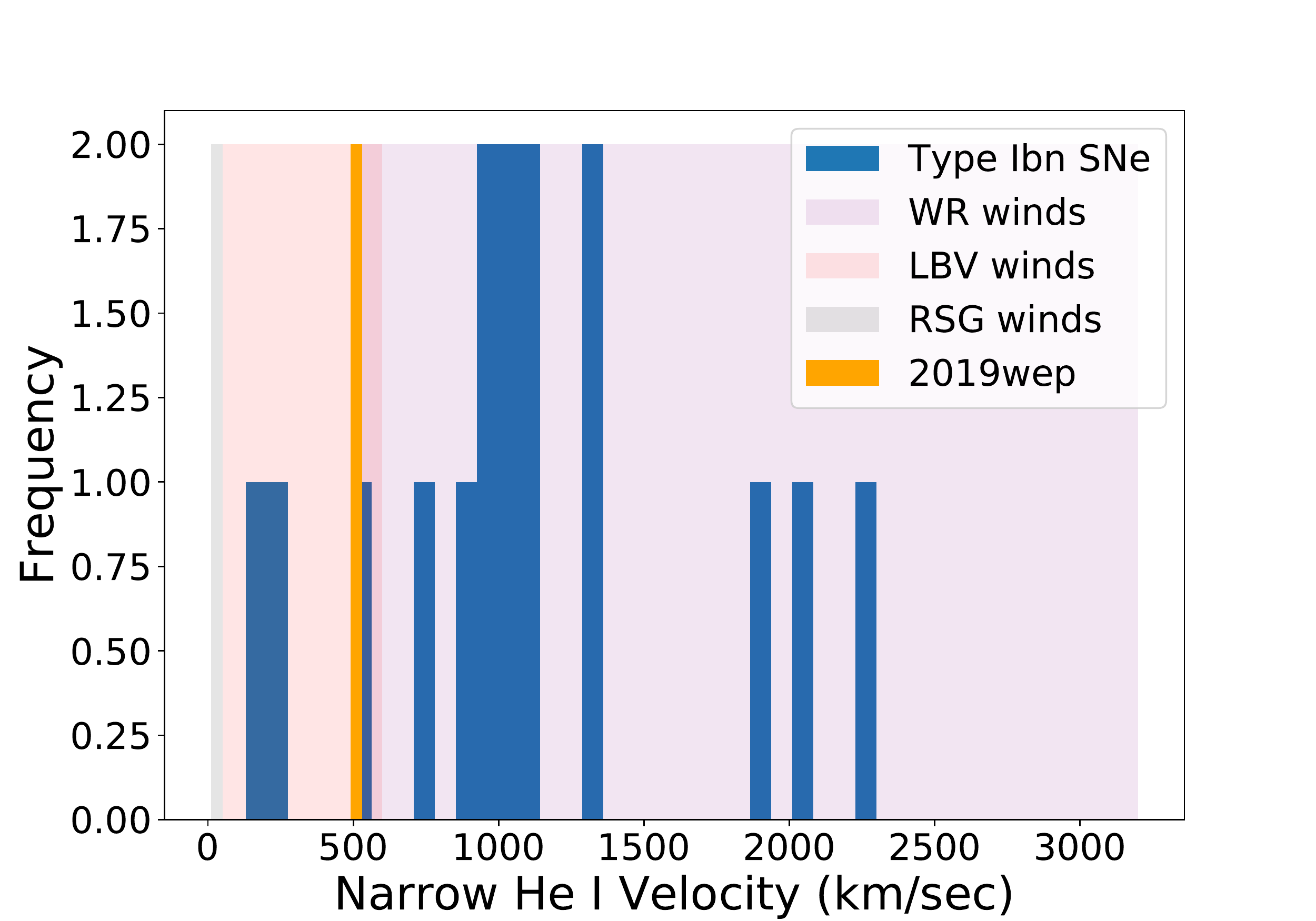}
	\end{center}
	\caption{Histogram representing the sample of SNe Ibn taken from \cite{2016MNRAS.456..853P} with blue bars. The orange shaded bar is the velocity with errors of SN 2019wep averaged over the first two days. The black, red and purple shaded regions represent the wind speeds of RSG, LBV and WR stars respectively. }
	\label{fig:histogram}
\end{figure}

The properties of stellar wind and CSM can be inferred probing the line emitting regions. Spectra of interacting SNe~IIn and SNe~Ibn are typically produced in different gas regions \citep{1985LNP...224..123C,1997Ap&SS.252..225C}. The emitting material moving at different velocities are indicated by different components of SNe with varying widths. The slow expanding, photoionised CSM is usually indicated by narrow emission lines (with velocities from a few hundreds to 6500 km sec$^{-1}$). This gas is unshocked CSM produced by the progenitor star before exploding as a SN. This gives information on the mass-loss history of the SN progenitors. When a clear P~cygni profile is identified, the position of the core of the blue-shifted absorption gives an idea of the expanding material. When this component is not detected, the velocity is estimated through the FWHM of the strongest \ion{He}{1} emission lines, obtained after deblending the full line profile with Gaussian fits. Figure~\ref{fig:histogram} shows the \ion{He}{1} velocity of SN 2019wep,   estimated from the weighted average of first two days to be 510 $\pm$ 20 km sec$^{-1}$. The expected range of WR wind speeds (500-3200 km sec$^{-1}$) \citep{2007A&G....48a..35C} is shown as the purple shaded region, the expected wind speeds of LBV stars (50-600 km sec$^{-1}$) \citep{2017hsn..book..403S} is shown in red  and the expected wind speed range of RSG stars (10-50 km sec$^{-1}$) \citep{2005A&A...438..273V} shown in black. Objects like SN~2019wep which show low CSM velocity are well- matched with other low velocity SNe~2011hw (200-250 km sec$^{-1}$) and 2005la (about 500 km sec$^{-1}$). For such SNe, an H$\alpha$ line of moderate strength is also found to be associated with the SNe. Scenarios of H lines along with modest stellar wind velocities (a few hundreds km sec$^{-1}$) are compatible with stellar progenitors which are transitioning between the LBV and the Wolf-Rayet (WNE-type) stages \citep{2008MNRAS.389..131P, 2015MNRAS.454.4293P, 2016MNRAS.456..853P}. PS1-12sk is one such event which showed narrow \ion{He}{1} lines and relatively weak intermediate-width H$\alpha$ lines and is uniquely associated with an elliptical galaxy. The plot also shows that the velocity of \ion{He}{1}, represented by horizontal bars, lies at the lower end of the velocity distribution of SNe~Ibn.

 The traditional description by \cite{1994PASP..106.1025H} defines LBVs as single stars between 60-100 M$_{\odot}$ at a transitional phase in the evolution of the most massive stars, between the main-sequence O-type stars and the H-deficient WR stars. However,  \cite{2015MNRAS.447..598S} found that the LBVs are surprisingly isolated from O-type stars.  \cite{2015MNRAS.447..598S} further claimed that many LBVs are likely to be the product of binary evolution and cannot drive the mass-loss mechanism leading to WR phase. \cite{2016ApJ...825...64H} again reverted the scenario proposed by  \cite{2015MNRAS.447..598S} highlighting their uneven sample and reconfirm that LBVs are evolved massive stars that have shed a lot of mass and are moving towards WR phase. 

Following the above description both single and binary progenitor scenario are equally likely in case of SN 2019wep with the progenitor having a reminiscent H$\alpha$ feature typical of LBV stars and gradually transitioning to a WR phase having a dense CSM owing to the mass-loss rates of the LBV.  It has been established that the sites of SNe Ibn are generally associated with star-forming environments \citep{2015MNRAS.449.1921P}, with one exceptional case where no/little association of star-formation activity was found (PS1-12sk, \citealt{2013ApJ...769...39S,2019ApJ...871L...9H}). Thus, the popular and speculative progenitors are typically single massive WR stars with mass $\geq$ 18 M$_{\odot}$ \citep{2008ApJ...687.1208T,2022arXiv220100955M} or a binary progenitor scenario which may be equally likely \citep{2007ApJ...657L.105F}.\\

\noindent
{\bf Origin of H$\alpha$ feature:}

The H$\alpha$ feature originating in SNe serve as an important classifier among the interacting SNe group. Although narrow \ion{He}{1} emission lines are frequently detected in the spectra of SNe~IIn, the relative strengths between H$\alpha$ and the most prominent optical lines \ion{He}{1} (5876 and 7065 \AA) determines the classification of the transient as a SN~IIn, SN~Ibn or even a transitional object between these two SN types. 

\begin{figure}
	\begin{center}
		%\vspace{-1.0cm}
		\includegraphics[width=\columnwidth]{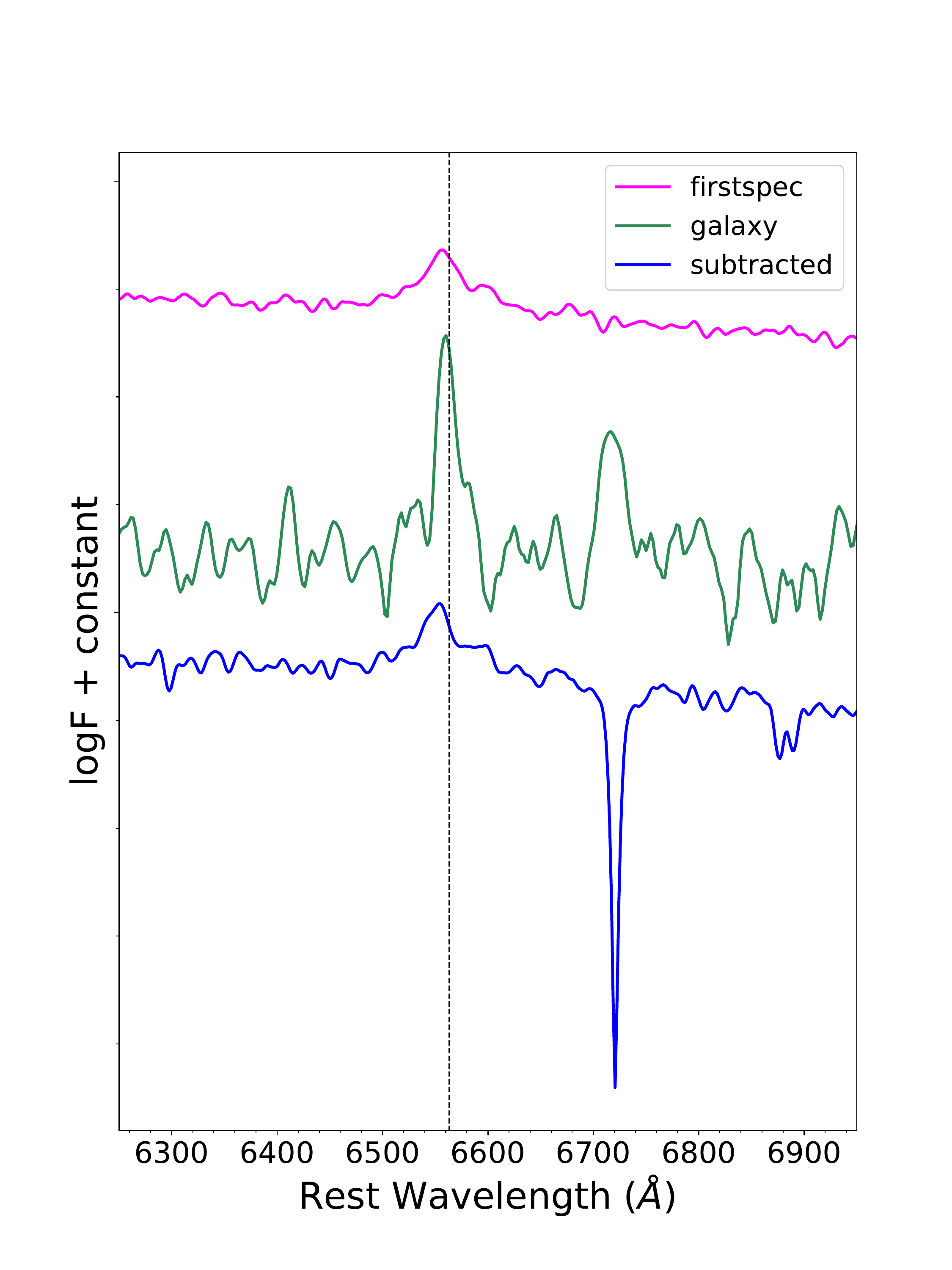}
	\end{center}
	\caption{The first day, galaxy and subtracted spectra of SN 2019wep taken with the LCO networks. A reminiscent H$\alpha$ is seen in the subtracted profile.}
	\label{fig:subtracted}
\end{figure}

In the two transitional SNe~IIn/Ibn 2005la and 2011hw, Balmer lines were prominently detected \citep{2008MNRAS.389..131P,2015MNRAS.454.4293P,2016MNRAS.456..853P}. The strength of H$\alpha$ feature in both these SNe are comparable with \ion{He}{1} lines. In the early spectrum of SN 2005la \citep{2014AJ....147...99M}, H$\alpha$ showed a narrow, marginally resolved (FWHM$\sim$400 km sec$^{-1}$) component, superposed on a broader base with a P~cygni profile while the highest resolution spectrum of SN 2011hw \citep{2015MNRAS.454.4293P} showed an unresolved H$\alpha$ (FWHM $\leq$ 250 km sec$^{-1}$) observed over a much broader wing and the intermediate H$\alpha$ component with FWHM monotonically increasing between 1350 km sec$^{-1}$ and 2350 km sec$^{-1}$. However, most of the SNe~Ibn show no prominent H$\alpha$ lines or, in some cases, the identification of H$\alpha$ is controversial. For example, the detection of multiple lines of \ion{C}{2} in the spectra of PS1-12sk \citep{2013ApJ...769...39S} argued - at least in the case of that object - against the identification of H lines.

\begin{figure}[h]
	\begin{center}
		%\vspace{-1.0cm}
		\includegraphics[width=\columnwidth]{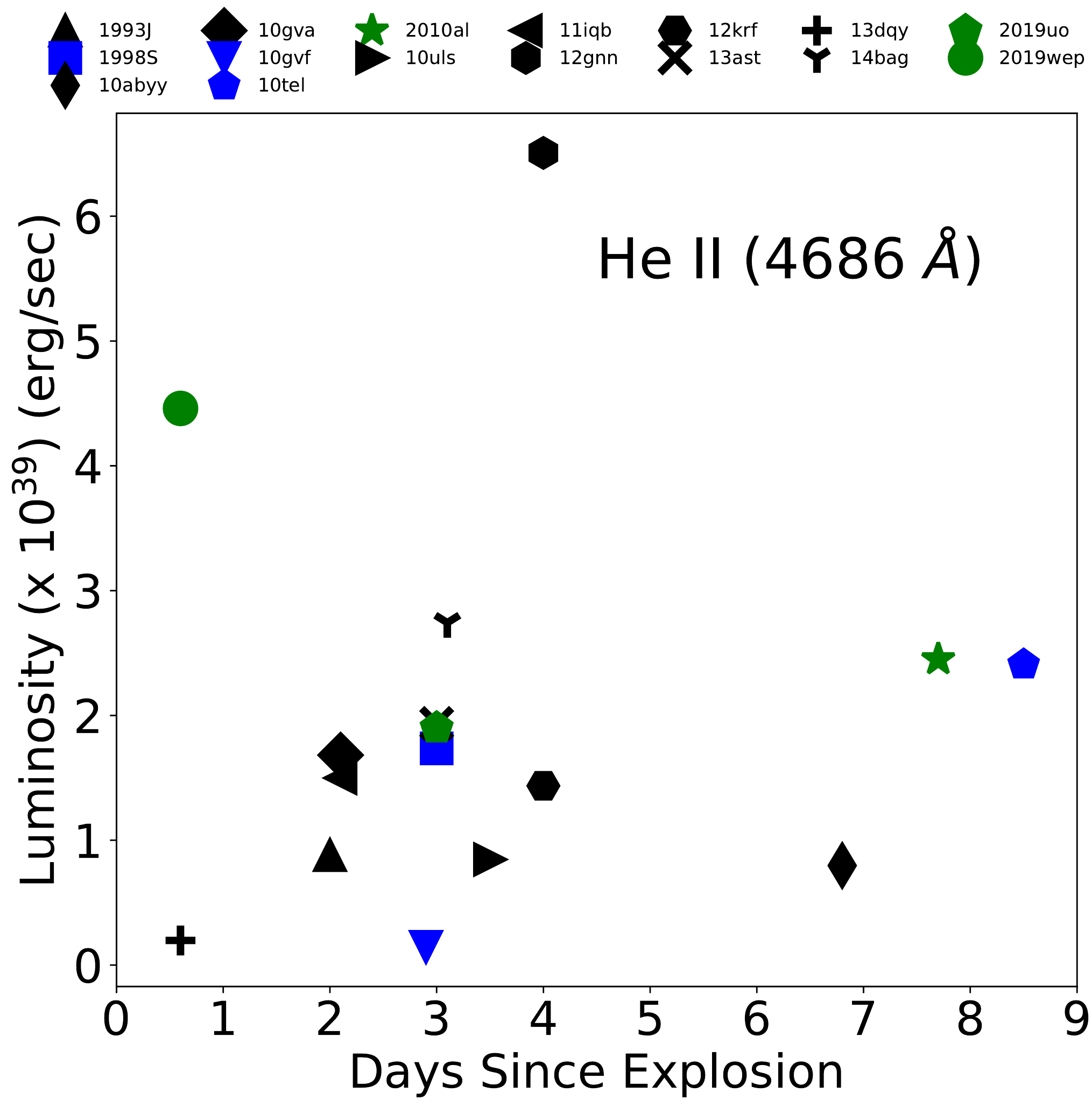}
	\end{center}
	\caption{ \ion{He}{2} 4686 \AA~ luminosity and equivalent width of a sample of SNe showing flash ionization features in their early spectra with SNe II in black, IIn in blue, and Ibn in green \citep{2016ApJ...818....3K, 2020ApJ...889..170G}. SN~2019wep is shown as a green circle.}
	\label{fig:flash_ionization.}
\end{figure}

We also see little H$\alpha$ at early time in SN~2019wep. To check whether the H$\alpha$ is from the host galaxy or is a reminiscent contribution from the SN, we obtained a spectrum in June 2021 from 2.0m FTN. This was used as a host galaxy template for subtraction on the first spectrum of the SN obtained with the same instrumental setup. The flux of the H$\alpha$ feature was matched. The subtracted spectrum is shown in Figure~\ref{fig:subtracted}. Even after spectral subtraction, we see a residual H$\alpha$ component with FWHM $\sim$ 1050 km sec$^{-1}$. When comparing with the first LCO spectrum, it looks like that one also had a broad H$\alpha$ base, not present in the host spectrum. So, we propose that at least some (and perhaps all) of the narrow H$\alpha$ seen is from the host, but we cannot rule out that some of it (especially because there's a broad base) is from the SN. We, however, concur that there may be a contamination due to \ion{C}{2}. In terms of spectroscopic properties, a transitioning behaviour like SN~2011hw is noticed for SN~2019wep.

To ascertain the origin of SNe~Ibn, a sample of 13 SNe~II from literature (including SNe~IIb, IIP and IIn; \cite{2016ApJ...818....3K}) and Ibn \citep{2017ApJ...836..158H, 2020ApJ...889..170G} , with flash ionisation signatures within 10 days of explosion, are collected. The H lines are usually contaminated by host galaxy lines, we therefore select the relatively isolated and unblended \ion{He}{2} 4686 \AA\ line. Since the \ion{He}{2} lines are much narrower than the lines from the SN ejecta, they can serve as a good tool to probe the flash-ionized CSM. While measuring the luminosities, we remove the continuum by fitting a linear function. Figure~\ref{fig:flash_ionization.} shows the typical luminosity of the \ion{He}{2} line in SN~2019wep and other SNe~II/IIn/Ibn. From the figure we see that only one SN along with SN~2019wep has been detected very early post explosion for which flash ionization signatures are detected owing to the recombination of CSM. The average luminosity of \ion{He}{2} for SN~2019wep is higher than almost all other SNe~II, IIn and Ibn which is indicative of the fact that the amount of CSM in SN~2019wep is larger than average SNe~II, IIn and Ibn sample at early time. This highlights the need for high cadence SN spectroscopy at very early phase to detect signatures of flash ionisation which are also indicative of presence of CSM and will in turn be useful to infer the possible progenitors. 
		
\section{Summary}
\label{5}

The paper summarises the photometric and spectroscopic evolution of the SN~Ibn 2019wep from $-0.4$ to $95$ days post maximum. The availability of early data points in TNS were useful to constrain the $V$-band maxima. The light curve decline is slower than the fast transients but is consistent with the SNe~Ibn group with a typical decay rate of $\sim$0.1~mag d$^{-1}$ \citep{2017ApJ...836..158H}.  Fast transients are one of the exciting findings with characteristic timescales of $\leq$ 10 days, missed in most of the traditional surveys either due to poor resolution of the spectrum or unavailability of follow-up spectroscopic observations. A recent study by \cite{2021arXiv210508811H} classified SNe~Ibn as a subclass of fast transients due to their very similar rise times and blue continuum. This study motivated us to compare the properties of SN~2019wep with samples of SNe~Ibn and fast transients. The limits on the absolute magnitude (M$_{V}$= -18.26 $\pm$ 0.20 mag and M$_{r}$ = -18.18 $\pm$ 0.95) shows that the SN lies at the fainter end of SN~Ibn subclass but is closer to the average absolute magnitude of the fast transients. 
 
The correlation plots also suggest that the rise time of SN~2019wep is higher than that of the fast transients but is shorter than the average SNe~Ibn value. Our analysis suggests that SN~2019wep decays fast and has low luminosity in comparison to other SNe~Ibn. We quote the results of the bolometric light curve modelling of SN~2019wep from \cite{2021arXiv211015370P}. The bolometric light curve of SN~2019wep has striking resemblance with SN~2019uo suggesting a comparable $^{56}$Ni mass in both SNe. The ejecta mass of SN~2019wep is consistent with SNe~Ib samples, even though the models cannot be directly comparable because of different physical mechanisms involved. The low $^{56}$Ni mass of SN~2019wep may suggest a significant fallback in the center which can also possibly explain the consistency of the ejecta mass of SN~2019wep with SNe~Ib sample. The color evolution of SN~2019wep is unique, placing it between SNe~Ib and SNe~Ibn. The spectroscopic features of SN~2019wep indicate that it is one of the rare SNe~Ibn with signatures of flash ionisation. The early, prominent flash ionisation lines of \ion{He}{2}, \ion{C}{3}, and \ion{N}{3} are detected in the spectra, similar to SN~2019uo and SN~2010al. The disappearance of narrow \ion{He}{1} features immediately after maximum and the transitioning to broad P~cygni \ion{He}{1} features hints towards a lateral shift from SN~Ibn to SN~Ib group. We interpret that the resemblance to SN~Ib is due to the fact that the SN ejecta that is likely hidden by the CSM interaction becomes evident when the recombination comes into play at late phases. The P~cygni subclass most likely originates from a He-rich shell around the progenitor surrounded by a dense CSM, or it may be due to viewing angle dependency. Asymmetricity of the CSM configuration also plays a major role in this aspect. This is also validated by the equivalent widths of \ion{He}{1} features. The estimated line velocities are lower than the average values of SNe~Ibn, but they show a faster evolution as compared to the ``emission" subclass. The low \ion{He}{1} velocity of CSM and the presence of residual H$\alpha$ in the spectroscopic features indicate that SN~2019wep is a transitional SN~Ibn with progenitor scenario lying between a LBV and a WR star. \cite{2022arXiv220100955M} essentially found that the progenitor of SNe~Ibn are WR stars with mass $\geq$ 18 M$_{\odot}$. While \cite{2021arXiv210508811H} estimate that the volumetric rate of SNe~Ibn population is 0.5$\%$ of CCSNe, \cite{2022arXiv220100955M} with their updated estimations found that the volumetric rates is 1$\%$ of CCSNe population. This is of course below the fraction of SNe~Ibn with masses $\geq$ 18 M$_{\odot}$ but many of them are undetected in optical because of emission of large fraction of UV radiation for these candidate SNe \citep{2022arXiv220100955M}. High cadence UV surveys are thus, quintessential to detect 'UV Ibn' SNe population.

\vspace{3cm}

\section*{Acknowledgments}
We thank the referee for several critical comments and useful suggestions which has improved the presentation of the paper. We thank the support of the staff of the Xinglong 2.16~m and Lijiang 2.4~m telescope, for their support during observations. This work was partially supported by the Open Project Program of the Key Laboratory of Optical Astronomy, National Astronomical Observatories, Chinese Academy of Sciences. Funding for the Lijiang 2.4~m telescope has been provided by Chinese Academy of Sciences and the People's Government of Yunnan Province. We acknowledge Weizmann Interactive Supernova data REPository (WISeREP; \url{http://wiserep.weizmann.ac.il}). DAH acknowledges support from NSF grant AST-1313404. The work of XW is supported by the National Science Foundation of China (NSFC grants  12033003, 11633002, and 11761141001), the Major State Basic Research Development Program (grant 2016YFA0400803), and the Scholar Program of Beijing Academy of Science and Technology (DZ:BS202002). This work makes use of data obtained with the LCO Network.  KM  acknowledges BRICS grant DST/IMRCD/BRICS/Pilotcall/ProFCheap/2017(G)  and the DST/JSPS grant, DST/INT/JSPS/P/281/2018 for the present work. IA is a CIFAR Azrieli Global Scholar in the Gravity and the Extreme Universe Program and acknowledges support from that program, from the European Research Council (ERC) under the European Union’s Horizon 2020 research and innovation program (grant agreement number 852097), from the Israel Science Foundation (grant number 2752/19), from the United States - Israel Binational Science Foundation (BSF), and from the Israeli Council for Higher Education Alon Fellowship. JZ is supported by the NSFC (grants 12173082, 11773067), by the Youth Innovation Promotion Association of the CAS (grants 2018081), and by the Ten Thousand Talents Program of Yunnan for Top-notch Young Talents. The LCO team is supported by NSF grants AST-1911151 and AST-1911225.
		

\software{lcogtsnpipe \citep{2011MNRAS.416.3138V, 2016MNRAS.459.3939V}, IRAF \citep{1986SPIE..627..733T, 1993ASPC...52..173T}, DAOPHOT \citep{1987PASP...99..191S}, HOTPANTS \citep{2015ascl.soft04004B}, superbol \citep{2016ascl.soft09019L}}

\bibliography{refag}{}
\bibliographystyle{aasjournal}



%\begin{landscape}

%\end{landscape}

\end{document}